\documentclass[smus]{snow2e}
\usepackage{graphics}
\makeatletter

\newcount\@tempcntc
\def\@citex[#1]#2{\if@filesw\immediate\write\@auxout{\string\citation{#2}}\fi
  \@tempcnta\z@\@tempcntb\m@ne\def\@citea{}\@cite{\@for\@citeb:=#2\do
    {\@ifundefined
       {b@\@citeb}{\@citeo\@tempcntb\m@ne\@citea\def\@citea{,}{\bf ?}\@warning
       {Citation `\@citeb' on page \thepage \space undefined}}%
    {\setbox\z@\hbox{\global\@tempcntc0\csname b@\@citeb\endcsname\relax}%
     \ifnum\@tempcntc=\z@ \@citeo\@tempcntb\m@ne
       \@citea\def\@citea{,}\hbox{\csname b@\@citeb\endcsname}%
     \else
      \advance\@tempcntb\@ne
      \ifnum\@tempcntb=\@tempcntc
      \else\advance\@tempcntb\m@ne\@citeo
      \@tempcnta\@tempcntc\@tempcntb\@tempcntc\fi\fi}}\@citeo}{#1}}
\def\@citeo{\ifnum\@tempcnta>\@tempcntb\else\@citea\def\@citea{,}%
  \ifnum\@tempcnta=\@tempcntb\the\@tempcnta\else
   {\advance\@tempcnta\@ne\ifnum\@tempcnta=\@tempcntb \else \def\@citea{--}\fi
    \advance\@tempcnta\m@ne\the\@tempcnta\@citea\the\@tempcntb}\fi\fi}
\makeatother

\def\lsim{\mathrel{\raise.3ex\hbox{$<$\kern-.75em\lower1ex\hbox{$\sim$}}}}
\def\gsim{\mathrel{\raise.3ex\hbox{$>$\kern-.75em\lower1ex\hbox{$\sim$}}}}
\def\lplm{\ell^+\ell^-}
\def\mpmm{\mu^+\mu^-}
\def\epem{e^+e^-}
\def\ca{%
   \def\.{\hskip-3pt plus 1pt}
   \def\vr{\vrule height 5pt depth 3pt}
   {\rlap{\raise 5pt \hbox{\vr}}\.\rightarrow\;}
}
\def\9{\hphantom0}
\def\ifmath#1{\relax\ifmmode #1\else $#1$\fi}
\def\ls#1{\ifmath{_{\lower1.5pt\hbox{$\scriptstyle #1$}}}}

\def\tanb{\tan\beta}

\def\hl{h^0}
\def\ha{A^0}
\def\hh{H^0}
\def\hpm{H^\pm}
\def\hsm{h^0_{\scriptscriptstyle\rm SM}}
\def\mhsm{m_{h^0_{\rm SM}}}
\def\mha{m_{\ha}}
\def\mhl{m_{\hl}}
\def\mhh{m_{\hh}}
\def\mhpm{m_{\hpm}}
\def\mz{m_Z}
\def\mw{m_W}
\def\mt{m_t}

\def\mzz{m_Z^2}
\def\mww{m_W^2}

\def\mpl{M_{\rm PL}}
\def\anti{\overline}

\def\fbi{~{\rm fb}^{-1}}

\def\gev{\,{\rm GeV}}

\def\br{{\rm BR}}
\def\rts{\sqrt s}
\def\gamhsm{\Gamma_{\hsm}^{\rm tot}}
\def\gam{\gamma}
\def\wstar{W^{\star}}

\onecolumn
\begin{document}

\vbox{  \large
\begin{flushright}
SCIPP 97/03 \\
UCD 97-06\\
hep--ph/9703391\\
\end{flushright}
\vskip1cm
\begin{center}
{\LARGE\bf
Weakly-Coupled Higgs Bosons and Precision Electroweak Physics}\\[1pc]

{\bf Howard E. Haber}\\ {\it Santa Cruz Institute for Particle
Physics, University of California, Santa Cruz, CA 95064}\\[2.mm]
{\bf Tao Han}\\ {\it Department of Physics,
University of California, Davis, CA 95616}\\[2.mm]
{\bf Frank S. Merritt}\\ {\it Department of Physics,
University of Chicago, IL 60637}\\[2.mm]
{\bf John Womersley}\\ {\it Fermi National Accelerator Laboratory, P.O.
Box 500, Batavia, IL 60510} \\[4.mm]
{\bf Precision Electroweak Physics Subgroup Conveners}\\
U.~Baur {\it (SUNY-Buffalo)} and M.~Demarteau {\it (Fermilab) }\\[3.mm]
{\bf Higgs Boson Discoveries Subgroup  Conveners}\\
C. Kao {\it (Univ. of Wisconsin)} and P.C. Rowson {\it (SLAC)}  \\[3.mm]
{\bf Higgs Boson Properties Subgroup  Conveners}\\
J. F. Gunion {\it (UC-Davis)}, R. Van Kooten {\it (Indiana Univ.)}
and L. Poggioli {\it (CERN)}\\

\vskip2cm
\thispagestyle{empty}

{\bf Abstract}\\[1pc]

\begin{minipage}{15cm}
We examine the prospects for discovering and elucidating the
weakly-coupled Higgs sector at future collider experiments.
The Higgs search consists of three phases: (i) discovery of a
Higgs candidate, (ii) verification of the Higgs interpretation of the
signal, and (iii) precision measurements of Higgs sector
properties.  The discovery of one Higgs boson with Standard Model
properties is not sufficient to expose the underlying structure of the
electroweak symmetry breaking dynamics.  It is critical to search
for evidence for a non-minimal Higgs sector and/or new physics
associated with electroweak symmetry breaking dynamics.  An improvement
in precision electroweak data at future colliders can play a useful
role in confirming the theoretical interpretation of the Higgs search
results.
\end{minipage}  \\
\vskip2cm
Summary Report \\ 
The Weakly-Coupled Higgs
Boson and Precision Electroweak Physics Working Group \\
1996 DPF/DPB Summer Study on New Directions for High Energy Physics\\
25 June--12 July, 1996, Snowmass, CO \\
\end{center}
}
\vfill
\clearpage

\twocolumn
\setcounter{page}{1}
\pagestyle{plain}
\title{Weakly-Coupled Higgs Bosons and Precision Electroweak Physics%
\thanks{This is the summary report of the Weakly-Coupled Higgs
Boson and Precision Electroweak Physics Working Group.
The full list of working group members can be found in the subgroup
reports that follow this summary report.  This work
was supported in part by the
U.S. Department of Energy and the National Science Foundation.}} 

\author{{\bf Howard E. Haber}\\ {\it Santa Cruz Institute for Particle
Physics, University of California, Santa Cruz, CA 95064}\\[2.mm]
{\bf Tao Han}\\ {\it Department of Physics,
University of California, Davis, CA 95616}\\[2.mm]
{\bf Frank S. Merritt}\\ {\it Department of Physics,
University of Chicago, IL 60637}\\[2.mm]
{\bf John Womersley}\\ {\it Fermi National Accelerator Laboratory, P.O.
Box 500, Batavia, IL 60510} \\[4.mm]
{\bf Precision Electroweak Physics Subgroup Conveners}\\
U.~Baur {\it (SUNY-Buffalo)} and M.~Demarteau {\it (Fermilab) }\\[3.mm]
{\bf Higgs Boson Discoveries Subgroup  Conveners}\\
C. Kao {\it (Univ. of Wisconsin)} and P.C. Rowson {\it (SLAC)}  \\[3.mm]
{\bf Higgs Boson Properties Subgroup  Conveners}\\
J. F. Gunion {\it (UC-Davis)}, R. Van Kooten {\it (Indiana Univ.)}
and L. Poggioli {\it (CERN)}\\
}
\maketitle


\begin{abstract}
We examine the prospects for discovering and elucidating the
weakly-coupled Higgs sector at future collider experiments.
The Higgs search consists of three phases: (i) discovery of a
Higgs candidate, (ii) verification of the Higgs interpretation of the
signal, and (iii) precision measurements of Higgs sector
properties.  The discovery of one Higgs boson with Standard Model
properties is not sufficient to expose the underlying structure of the
electroweak symmetry breaking dynamics.  It is critical to search
for evidence for a non-minimal Higgs sector and/or new physics
associated with electroweak symmetry breaking dynamics.  An improvement
in precision electroweak data at future colliders can play a useful
role in confirming the theoretical interpretation of the Higgs search
results.
\end{abstract}

\section{INTRODUCTION}

Present day colliders test the Standard Model at an energy scale of
order 100~GeV.  Precision experiments at LEP, SLC and Tevatron
(with some additional measurements at lower energies) have
measured more than twenty separate experimental observables, and have
confirmed the Standard Model predictions with an accuracy of one part in
a thousand \cite{PRECISION,LEPEWWG}.
A few anomalies in the data could suggest hints of
new physics beyond the Standard Model \cite{NEWPHENO},
although no deviations have been rigorously confirmed.

Nevertheless, the verification of the Standard Model is not yet
complete.  Absent to date is any experimental signal that sheds light
on the dynamics responsible for electroweak symmetry breaking.  Any
consistent theory of electroweak symmetry breaking must generate
Goldstone bosons
which are absorbed by the $W^\pm$ and $Z$ gauge bosons, thereby
generating the gauge boson masses.
The Standard Model posits that electroweak symmetry breaking is
due to the dynamics of a weakly-coupled complex
doublet (with hypercharge one) of elementary scalar fields.  The
physical consequence of this model is the existence of a CP-even neutral
Higgs boson with mass roughly of order $m_Z$.
Extensions of this model can
easily be constructed, in which the scalar sector is enlarged.  The
resulting model then contains a {\it non-minimal} Higgs sector
consisting of neutral Higgs bosons (of definite or indefinite CP
depending on the model) and charged Higgs bosons~\cite{HUNTERS}.

The best motivated non-minimal Higgs sector is the two Higgs doublet
model.  Starting with two complex scalar doublets of hypercharge $\pm 1$
respectively, one finds a Higgs sector (after three Goldstone bosons are
absorbed to give mass to the $W^\pm$ and $Z$) consisting of five states:
a light CP-even Higgs scalar, $\hl$, a heavy CP-even scalar, $\hh$, a
CP-odd scalar $\ha$, and a charged Higgs pair, $H^\pm$.  This is the Higgs
sector of the minimal supersymmetric extension of the Standard
Model (MSSM)~\cite{HUNTERS,SUSY,SUSYGROUP}.

In the
global fits of LEP, SLC, and Tevatron data based on the Standard Model,
there is weak (but non-trivial) sensitivity to the Higgs boson mass by
virtue of Higgs mediated radiative corrections.  The most recent global
fits find that $\mhl<550$~GeV at $95\%$ CL \cite{LEPEWWG}, although some
care needs to be taken in interpreting this limit \cite{rosner}.
The potential for improving this bound at future colliders is discussed
in Section II.
In the context of the Standard Model, the Higgs boson in this mass range
is necessarily weakly-coupled.
Moreover, such fits also apply to non-minimal Higgs
sectors in which the lightest Higgs scalar ($\hl$) is separated in mass
from heavier non-minimal Higgs states.  Therefore, there is a strong
motivation to conduct a vigorous experimental search for
weakly-coupled Higgs bosons at LEP and future colliders.

If a Higgs boson with Standard Model properties were discovered, then
one might naively conclude that the search for the model of the
elementary particles has been completed.  However, theorists strongly
believe that the Standard Model cannot be the fundamental model of
particles.  Apart from the many parameters of the Standard Model which
must be inserted by hand (with no explanation), there is a theoretical
problem in the Standard Model associated with the
very large hierarchy of energy scales.  We know that
the Planck scale, $\mpl\simeq 10^{19}$ GeV, exists in nature; it
characterizes the energy scale above which gravitational interactions
cannot be neglected relative to the strong and electroweak interactions
of the elementary particles.  Given the existence of such a large
energy scale, one must explain how the scale of electroweak symmetry
breaking, which is so small when expressed in units of the Planck scale
($m_Z\simeq 10^{-17}\mpl$), could be generated by a fundamental theory
of particles that includes gravity.
Related to this question is the theoretical
problem of generating a ``naturally'' light Higgs boson (with a mass of order
$\mz \ll \mpl$), since in the Standard Model, there is no symmetry
that can protect the mass of an elementary scalar from being driven up
to $\mpl$ via radiative corrections.  These problems are
intimately connected with the dynamics that generates
electroweak symmetry breaking.

Attempts to solve the problem of hierarchy and the related problem of
the unnaturally light Higgs boson inevitably lead to the existence of
new physics at the 1~TeV energy scale or below. Possible mechanisms invoke
either supersymmetry \cite{SUSY} (a symmetry that can protect
the masses of elementary scalars) or dynamical electroweak symmetry
breaking \cite{TC} (which typically eliminates elementary
scalar fields completely).
Which path nature chooses can only be determined through
experimentation.  Thus the central goals of the future colliders program
are: to explore the dynamics of electroweak symmetry breaking, and
to determine its implications for the structure of the Standard Model and
the nature of physics that lies beyond the Standard Model.

In this report, we assume that
nature chooses a weakly-coupled Higgs sector as the source of
electroweak symmetry breaking dynamics.
We do not address the
alternative approach which invokes strong interaction dynamics as
the source of electroweak symmetry breaking.
The phenomenology of the electroweak symmetry breaking sector
in this latter case is explored by the
Strongly Interacting Electroweak Symmetry Breaking Working Group
\cite{SEWS}.
The focus of this working group is the weakly-coupled Higgs sector of
the Standard Model, and possible non-minimal Higgs sector extensions
(including the Higgs sector of the MSSM).
Although there is considerable freedom for the structure of
the scalar sector (even after imposing all known theoretical and
phenomenological constraints), models of the scalar sector
often exhibit the following structure: (i) the lightest scalar
($\hl$) is a CP-even neutral Higgs boson with couplings closely
approximating those of the Standard Model Higgs boson ($\hsm$),
and (ii) additional Higgs scalars (neutral Higgs bosons with
definite or indefinite CP quantum numbers and charged Higgs bosons) are
expected to be heavier (perhaps significantly heavier) than $\hl$,
although still weakly-coupled.  This is the so-called {\it
decoupling limit} which will be discussed in Section IIIA.\footnote{It
could be that all scalar states are somewhat close in mass, with no
state possessing couplings that match those predicted by
the Standard Model. This case is actually simpler
to address experimentally and interpret theoretically.}   In this case,
the discovery of $\hl\simeq\hsm$ is not sufficient to probe the
underlying structure of the electroweak symmetry breaking sector.  The
essence of the decoupling limit is that the existence of a light CP-even
Higgs boson with properties closely approximating those of
$\hsm$ is consistent with many possible non-minimal Higgs sectors.
Thus, the
discovery of the heavy non-minimal Higgs scalars is essential in order
to probe the details of the electroweak symmetry breaking dynamics.

The MSSM provides a natural framework for light elementary Higgs
scalars.  The Higgs sector of the MSSM is a constrained
two-Higgs-doublet model, whose tree-level properties are determined by
two free parameter (typically chosen to be the mass of the CP-odd state,
$\ha$, and the ratio of vacuum expectation values, $\tanb$).  The
decoupling limit of the model corresponds to $\mha\gg\mz$; in this
limit, the properties of $\hl$ become identical to those of $\hsm$.
Extensions of the MSSM Higgs sector are also possible.
For example, the simplest non-minimal supersymmetric extension of the
Standard Model (NMSSM) consists of a Higgs sector with two doublets and
one singlet of complex Higgs fields \cite{nmssm}.
Thus, a detailed exploration
of the scalar sector has the potential for probing both the electroweak
symmetry breaking dynamics and the underlying supersymmetric structure
of the theory.

Present experimental data tells us that the Higgs sector must be
compatible with
\newcounter{mynum}
\begin{list}{(\roman{mynum})}{\usecounter{mynum} \itemsep 0in
\labelwidth .5in \leftmargin .7in}
\item $\rho\equiv\mw^2/\mz^2\cos^2\theta_W\simeq 1$;
\item the absence of significant Higgs mediated flavor changing neutral
currents;
\item the absence of significant virtual charged Higgs
mediated effects (which can contribute, {\it e.g.}, to $B^0$--$\overline
{B^0}$ mixing, $b\to s\gamma$ and $Z\to b\bar b$).
\end{list}
Even after imposing
such model constraints, there is still significant freedom in the
structure of the Higgs sector.  Exotic Higgs sectors (beyond those
mentioned above) are easily constructed that satisfy all present day
phenomenological constraints.  Such Higgs sectors could arise in models
with extended gauge groups, models with exotic scalar multiplets, or models
with a lepton number violating sector ({\it e.g.}, in R-parity violating
models of low-energy supersymmetry \cite{rparity},
in which there is no distinction
between scalar lepton superpartners and Higgs bosons).
Sorting out the details of the scalar sector will be one of the
fundamental challenges for future collider experimentation.

The Weakly-Coupled Higgs Boson and Precision Electroweak Physics Working
Group program consisted of the following tasks:

\begin{enumerate}
\item Extend present day precision tests of the Standard Model

\noindent This will
serve to tighten constraints on the Higgs sector and perhaps uncover
deviations from the Standard Model and provide evidence for new physics
beyond the Standard Model.

\item Evaluate the Higgs boson discovery reach of future colliders

\noindent
High energy colliders are needed to directly produce the massive
Higgs bosons. However, the cleanest decay channels of the Higgs boson
usually have rather small branching ratios. Thus, high luminosity is
critical to insure that Higgs signals can be extracted from the Standard
model backgrounds.  In Table~\ref{colliders}, we list the approved
and possible future collider facilities considered in
our study.  Furthermore, special features of the collider
detectors (such as the high resolution for the electromagnetic
calorimeter for $\hsm \to \gamma \gamma$, and high $b$ tagging
efficiency) are also required in order to maximize the
significance of the Higgs signal.  Thus,
establishing the discovery reach for future colliders is
an important and non-trivial first step in the pursuit of the Higgs
boson.


\item Consider precision measurements of $\hl$ properties

\noindent
The discovery of the Higgs boson will complete the
experimental verification of the Standard Model.
Once the Higgs boson is discovered, one must check that it does
indeed possess couplings to particles proportional to their masses.
One should quickly be able to verify that the properties of the scalar
state roughly match those expected for the Standard Model Higgs boson.
More precise measurements may be required to detect deviations
of the observed Higgs properties from that of the $\hsm$.
The difficulty of this latter task will
depend on how close one is to the decoupling limit
(see Section IIIA).

\item  Evaluate the potential for direct detection of the non-minimal
Higgs states and the measurement their properties

\noindent
This is essential for probing the nature of the electroweak symmetry
breaking dynamics.  In addition, the non-minimal Higgs states may be
sensitive to physics associated with the hierarchy problem (for example,
the properties of the non-minimal Higgs states in supersymmetric models
can provide important checks of the supersymmetric dynamics).  The
non-minimal Higgs sector imposes the most stringent requirements on the
collider facility.  To accomplish this task may require the highest
energies and luminosities now being considered.


\end{enumerate}

\begin{table}[htb]
\caption{Approved and possible future collider facilities
considered in this study.  LEP-2 is currently running, but has not yet
reached its design energy and luminosity \protect \cite{lepdesign}.
Experimentation at the Tevatron Main Injector
(M.I.) is often referred to in the text as Run II.
Center of mass energy $\protect\sqrt{s}$
and design annual integrated luminosity are specified.}
\label{colliders}
\setlength{\tabcolsep}{6pt}
\renewcommand\arraystretch{1.2}
\begin{center}
\begin{tabular}{lccc}
\hline\hline
\multicolumn{1}{c}
{Name}& Type& $\sqrt s$& Annual $\int{\cal L}$\\[2pt]
\hline
{\em Approved:}&&& \\
LEP-2&    $e^+e^-$&  192 GeV&  170 pb$^{-1}$     \\
Tevatron (M.I.)& $p\bar p$& 2 TeV& 2 fb$^{-1}$ \\
LHC&       $pp$&  14 TeV&  100 fb$^{-1}$  \\[2pt]
\hline
{{\em Possible:}}&&&     \\
TeV-33& $p\bar p$& 2 TeV& 30 fb$^{-1}$ \\
NLC$^\dagger$ & $e^+e^-$ & 0.5--1.5 TeV & 50--200 fb$^{-1}$ \\
\multicolumn{4}{l}
{\qquad ($^\dagger$ with $e\gamma,\gamma\gamma,e^-e^-$
options)} \\
FMC& $\mu^+\mu^-$& 0.5--4 TeV&  50--1000 fb$^{-1}$ \\[2pt]
\hline \hline
\end{tabular}
\end{center}
\end{table}

This report consists of four parts.
Following this Introduction, Section II briefly summarizes the
results of the Precision electroweak physics subgroup. Section III
discusses some theoretical issues that are important for the
considerations of the Higgs discovery and properties subgroups.
Section IV summarizes the essentials of Higgs
phenomenology at future colliders.  The
conclusions and some final thoughts are given in Section V.
The details underlying Sections II and IV can be found in the subgroup
reports that follow this summary report \cite{PRECGROUP,gunreport}.

\section{PRECISION ELECTROWEAK PHYSICS AT FUTURE COLLIDERS}

In the electroweak Standard Model, there are two coupling parameters,
$g$ and $g'$, of $SU(2)_L\times U(1)_Y$ gauge interactions. The vacuum
expectation value of the scalar field, $v$, sets the mass scale.
At tree level, the $W^\pm$ and $Z$ boson masses $\mw, \mz$,
as well as the weak mixing angle $\sin\theta_W$, are determined
by these three parameters. Alternatively, one may use the precisely
measured quantities---the electromagnetic coupling constant
$\alpha$, the muon decay constant $G_\mu$ and $m_Z$---as inputs
to evaluate the other electroweak parameters. When the radiative
corrections are taken into account, the relations among
these parameters become dependent on $m_t$, $\mhsm$ as well as
other possible contributions from new physics. Therefore,
precision electroweak measurements not only check the consistency
of the Standard Model, but also constrain $\mhsm$ and other new physics
\cite{PT,LangAlt}.

The Precision Electroweak Physics subgroup \cite{PRECGROUP}
paid special attention to the measurements of $m_W$, $m_t$
and $\sin\theta_W$ at future collider
experiments. The implications for the constraints on $\mhsm$ are also
discussed.

Currently, the world average values for $\mw$ and $m_t$ are
\begin{equation}
m_W=80.356\pm 0.125~{\rm GeV}, \quad
m_t=175\pm 6~{\rm GeV}.
\label{mwdirect}
\end{equation}

The precision which can be achieved for $\mw$ and $m_t$ measurements
at different colliders is summarized in Table~\ref{TAB:WMASS} and
Table~\ref{TAB:TOP}, respectively.  Table entries are taken from
Ref.~\cite{PRECGROUP} unless otherwise indicated.

%
\begin{table}[th]
\renewcommand\arraystretch{1.2}
\setlength{\tabcolsep}{1pc}
\begin{center}
\caption{Expected $W$ mass precision at future colliders.}
\label{TAB:WMASS}
\begin{tabular}{lc}
\hline
\hline
\noalign{\vskip3pt}
Collider & $\delta \mw$ (MeV)\\[3pt]
\hline
\noalign{\vskip3pt}
NuTeV \protect\cite{CCFR} &  100 \\
HERA (1000~pb$^{-1}$) & \960 \\
LEP-2 (4$\times 25$~pb$^{-1}$) &  144 \\
LEP-2 (4$\times 500$~pb$^{-1}$) &\940 \\
Tevatron (2~fb$^{-1}$) &\935 \\
TeV-33 (10~fb$^{-1}$) & \920 \\
LHC (10~fb$^{-1}$)  & \915 \\[1.mm]
NLC (50~fb$^{-1}$)~\protect\cite{ecfa} & \915 \\
FMC (10~fb$^{-1}$)~\protect\cite{muonmwmt} & \920 \\[1.mm]
\hline\hline
\end{tabular}
\end{center}
\end{table}
\begin{table}[th]
\begin{center}
\caption{Expected top quark mass precision at future colliders.}
\label{TAB:TOP}
\setlength{\tabcolsep}{1.5pc}
\begin{tabular}{lc}
\hline
\hline
\noalign{\vskip3pt}
Collider & $\delta m_t$ (GeV) \\[3pt]
\hline
\noalign{\vskip3pt}
Tevatron (2~fb$^{-1}$) &  4 \\
TeV-33 (10~fb$^{-1}$) & 2 \\
LHC (10~fb$^{-1}$)  &   2 \\
NLC (50~fb$^{-1}$)~\protect\cite{ecfa} & 0.12 \\
FMC (10~fb$^{-1}$)~\protect\cite{muonmwmt} & 0.2 \\[1.mm]
\hline
\hline
\end{tabular}
\end{center}
\end{table}

The weak mixing angle is conveniently defined by
\begin{equation}
\sin^2\theta^{\it lept}_{\it eff} ={1\over 4}\left (1-{g_{V\ell}\over
g_{A\ell}}\right )\, ,
\end{equation}
where $g_{V\ell}$ and $g_{A\ell}$ are the effective vector and axial vector
coupling constants of the leptons to the $Z$ boson.
They are measured with very high precision from $Z$ leptonic
decays \cite{PRECISION,LEPEWWG} at LEP-I (forward-backward asymmetries)
and at SLC (left-right asymmetries).  The relation between
$\sin^2\theta^{\it lept}_{\it eff}$ and
the weak mixing angle in the $\overline{\rm MS}$ scheme,
$\sin^2\hat\theta_W(M_Z)$ is given by
\begin{equation}
\sin^2\theta^{\it lept}_{\it eff}\simeq\sin^2\hat\theta_W(M_Z)+0.00028.
\end{equation}
A fit to the combined current LEP-I and SLD asymmetry data
yields \cite{LEPEWWG}
\begin{equation}
\sin^2\theta^{\it lept}_{\it eff}=0.23165\pm 0.00024.
\label{EQ:ASYM}
\end{equation}

The anticipation for the measurement on $\sin^2\theta^{\it lept}_{\it eff}$
at future experiments is summarized in Table~\ref{TAB:SINW}.
\begin{table}[th]
\begin{center}
\caption{Anticipated precision for $\sin^2\theta^{\it lept}_{\it eff}$
measurement at future colliders.}
\label{TAB:SINW}
\setlength{\tabcolsep}{1pc}
\begin{tabular}{lc}
\hline  \hline
\noalign{\vskip3pt}
Collider & $\delta \sin^2\theta^{\it lept}_{\it eff}\
                  \left(\times 10^{-4}\right)$ \\[5pt]
\hline
\noalign{\vskip3pt}
SLC2000 \protect\cite{SLC2000}&  1.2 \\
TeV-33 (10~fb$^{-1}$) & 2 \\
LHC (10~fb$^{-1}$)  &  3 \\
NLC (10~fb$^{-1}$) & 0.6 \\[1.mm]
\hline
\hline
\end{tabular}
\end{center}
\end{table}

\begin{figure}[h]
\leavevmode
\begin{center}
\resizebox{8.5cm}{!}{%
\includegraphics{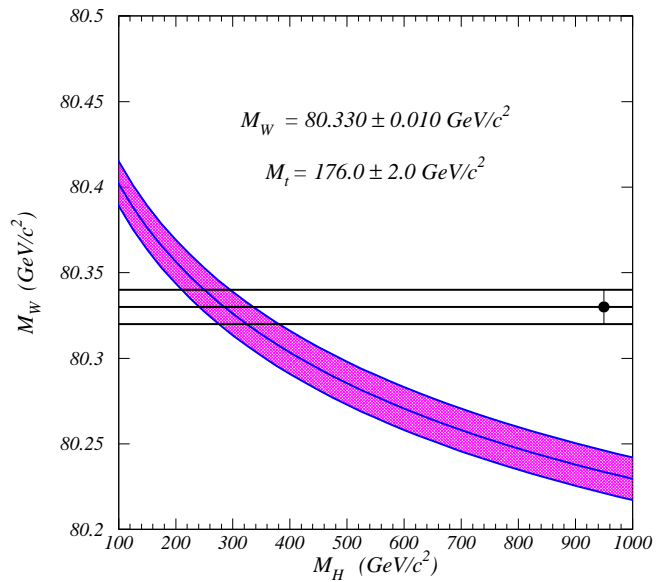}}
\end{center}
\caption{$\mw$ versus $\mhsm$ for $m_t=176\pm
2$~GeV/c$^2$. The theoretical predictions incorporate the effects of
higher order electroweak and QCD corrections.}
\label{FIG:mwmh}
\end{figure}

The high precision measurements of $\alpha$, $G_\mu$ and $\mz$,
along with the improved measurements of $\mw$, $m_t$ and
$\sin\theta_W$, may indirect shed light on the Standard Model Higgs
boson mass $\mhsm$. As an illustration, Fig.~\ref{FIG:mwmh} shows the
mass correlation for $\mw$ versus $\mhsm$ with $\mt= 176 \pm 2$ GeV.
A measurement of the $\mw$ with a precision of $\delta
\mw =10$~MeV and of $\mt$ with an accuracy of 2~GeV
thus translates into an indirect determination of the Higgs boson mass
with a relative error of about
\begin{equation}
\delta \mhsm/\mhsm \approx 20\%.
\end{equation}
However, it should be noted that to reach such a high precision,
other sources of uncertainty, such as $\alpha(\mz^2)$, $\alpha_s$
and theoretical uncertainties that arise when extracting $\mw$
\cite{berndk} and $\mt$ \cite{scottw}, must be kept under control.

\section{THEORETICAL CONSIDERATIONS FOR THE WEAKLY-COUPLED HIGGS SECTOR}

\subsection{The Decoupling Limit}

In this section, we discuss the theoretical implications of the
discovery of the first neutral Higgs boson, denoted by $\hl$.  Once this
state is discovered, one must check its theoretical interpretation.  A
Higgs state is predicted to couple to particles with coupling
strengths proportional to the particle masses.  After its initial
discovery, it should be straightforward to verify whether its properties
roughly match those expected of the Standard Model Higgs boson, $\hsm$.

In order to interpret the significance of the first Higgs discovery, it
is important to appreciate the concept of the {\it decoupling limit}
\cite{habernir,DECP}.
First, consider the Standard Model Higgs boson.  At tree-level, the
Higgs self-coupling is related to its mass.  If $\lambda$ is the quartic
Higgs self-interaction strength, then $\lambda= 3\mhsm^2/v^2$
(where $v\simeq 246$~GeV is the Higgs vacuum expectation value
which is fixed by the $W^\pm$ mass: $v=2\mw/g$).  This
means that one cannot take $\mhsm$ arbitrarily large without the
attendant growth in $\lambda$.  That is, the heavy Higgs limit in the
Standard Model is non-decoupling.  In models of a non-minimal Higgs
sector, the situation is more complex. In some models
(with the Standard Model as one example), it is not
possible to take any Higgs mass much larger than ${\cal O}(v)$ without
finding at least one strong Higgs self-coupling \cite{habernir}.
In other models, one
finds that the non-minimal Higgs boson masses can be taken large at
fixed Higgs self-couplings.
Such behavior can arise if the model possesses one (or more)
additional independent mass parameters beyond the diagonal scalar
squared-masses. In the limit where the additional mass parameters
are taken large [keeping the dimensionless
Higgs self-couplings fixed and $\lsim{\cal O}(1)$], the heavy Higgs
states decouple, while both light and heavy Higgs bosons remain
weakly-coupled.  In this {\it decoupling limit},
exactly one neutral CP-even Higgs
scalar remains light, and its properties are precisely those of the
(weakly-coupled) Standard Model Higgs boson.

In this report, we shall always assume that all Higgs scalars are
weakly-coupled (hence the name of this working group). Then, the
decoupling limit is one where $\hl\simeq\hsm$,
$\mhl\simeq{\cal O}(\mz)$, and all other non-minimal Higgs states are
significantly heavier than $\mhl$.  Squared-mass splittings of the heavy
Higgs
states are of ${\cal O}(\mz^2)$, which means that all heavy Higgs states
are approximately degenerate, with mass differences of order
$\mz^2/\mha$ (here $\mha$ is approximately equal to the common heavy
Higgs mass scale).  In contrast, if the non-minimal Higgs sector is
weakly coupled but far from the decoupling limit, then $\hl$ is not
separated in mass from the other Higgs states.  In this case, the
properties\footnote{The basic property of the Higgs coupling strength
proportional
to mass is maintained.  But, the precise coupling strength patterns of
$\hl$ will differ from those of $\hsm$ in the non-decoupling limit.}
of $\hl$ differ significantly from those of $\hsm$.

The decoupling limit arises naturally in many approaches.
For example, in models of Higgs
doublets (and singlets), with no artificial discrete symmetries imposed,
the decoupling limit is reached when off-diagonal Higgs mass parameters
are taken large.  Naturalness properties suggest that all such
parameters should reflect the highest possible energy scale consistent
with the model.  In models that introduce new TeV scale physics (to
explain the dynamics of electroweak symmetry breaking), these mass
parameters are expected to be associated with this new physics.  The
paradigm
for this discussion is the MSSM.  In the MSSM, the decoupling regime is
reached once $\mha\gsim 2\mz$.  The parameter $\mha$ arises in the MSSM
from the supersymmetry breaking sector.  The success of the Standard
Model in accounting for precision electroweak data suggests that if the
MSSM is correct, then the supersymmetry breaking scale is somewhat
higher than $\mz$ (though it must not be much higher than 1~TeV if it is
to explain the origin of the electroweak scale).  Likewise, one
might expect $\mha$ to also be somewhat higher than $\mz$.
Thus, in the MSSM, there is some expectation that the Higgs sector
approximately satisfies the decoupling limit.  Although
this argument is clearly not definitive, it will become more
persuasive if
supersymmetry and/or the non-minimal Higgs sector is not discovered at
LEP-2 or the Tevatron.

The phenomenological consequences of the decoupling regime are both
disappointing and challenging.  In this case, $\hl$ (once discovered)
will exhibit all the expected properties of $\hsm$.  The existence of
the non-minimal Higgs sector will still be unconfirmed.  It will require
precision measurements or the highest energies and luminosities at
future colliders to either
detect a deviation from Standard Model Higgs physics or to directly
detect the non-minimal Higgs states and explore their properties.  In
contrast, in the non-decoupling regime, more than one Higgs state is
expected to populate the mass region where the first Higgs boson is
found.  The properties of the first Higgs state will show a marked
deviation from $\hsm$ properties.  Experiments that can discover the
Higgs boson will have access to many scalar sector observables.

\subsection{Implications of a Higgs Discovery \\ for New High Energy
Scales}

Phenomenologists and experimentalists who plan the Higgs searches at
future colliders spend much effort in designing a search for the
Standard Model Higgs boson.  However, the term ``Standard Model Higgs
boson''
is meaningless unless additional information is provided.  This
is because the Standard Model itself cannot be a fundamental theory of
particle interactions.  It must break down once the energy is raised
beyond some critical scale $\Lambda$.  What is the value of $\Lambda$?
Of course, this is unknown at present.  $\Lambda$ can lie anywhere
between a few hundred GeV and the Planck scale ($M_{\rm PL}\simeq
10^{19}$~GeV).

Theorists who study the phenomenology of the Standard Model usually
do not need to know the value of $\Lambda$.  At energy scales below
$\Lambda$, the new physics beyond the Standard Model decouples,
leaving a low-energy effective theory which looks almost exactly like
the Standard Model.  However, the discovery of the Higgs boson provides
an opportunity to probe the value of $\Lambda$.  Consider the behavior
of the quartic
Higgs self coupling, $\lambda$, as a function of the energy scale.  At
low-energies, $\lambda= 3\mhsm^2/v^2$.  If one solves the one-loop
renormalization group equation for $\lambda(\mu)$,  one finds that
$\lambda$ increases with energy scale, $\mu$.  Eventually
$\lambda(\mu)$ becomes infinite at the so-called Landau pole.
Although this behavior could have been
an artifact of the one-loop approximation,
lattice results confirm that the theory breaks down at scales near the
Landau pole \cite{lattice}.  That is, we may associate $\Lambda$ with
the Landau pole.
Conversely, fixing the value of $\Lambda$ leads to an upper bound on
the low-energy value of $\lambda$, or equivalently to an upper bound on
$\mhsm$.  For example, if $\Lambda=\mpl$, then
$\mhsm\lsim 200$~GeV \cite{cabibbo,schrempp}.
Lower values of $\Lambda$ imply a higher Higgs mass upper bound.  Since
$\Lambda$ had better be larger than $\mhsm$ (since we are assuming the
Standard Model is a valid low-energy effective theory over some range of
energies), one can deduce an absolute Higgs mass upper bound of about
700--800~GeV. Similar conclusions are reached by lattice
computations \cite{lattice}.

The stability of the Higgs potential also
places non-trivial constraints on the Higgs mass, due to the large
value of the top quark mass.  (More refined limits require only a
metastable potential with a lifetime that is long
compared to the age of the universe.)  For example, recent computations
of Refs.~\cite{altarelli} and \cite{quiros} show that if
$\Lambda=M_{\rm PL}$, then for $m_t=175$~GeV the Higgs mass must be
larger than about 120~GeV.  If a Higgs boson were discovered whose mass
lies below this limit, then one would conclude that new physics beyond
the Standard Model must exist at some scale below $\mpl$.  As an
example,
if a Higgs boson of mass 100~GeV were discovered, then new physics
beyond the Standard Model must enter at or below an energy scale of
order $\Lambda=1000$~TeV (based on the graphs presented in
Ref.~\cite{quiros}).  Of course, in this case, if all the new physics
were
confined to lie in the vicinity of 1000~TeV, then LHC phenomenology would
find no deviations from the Standard Model.  Thus, physicists who
plan searches for the Standard Model Higgs boson are not wasting their
time.  In particular, even if $\Lambda$ is rather close to the TeV
scale, one would expect the lightest Higgs boson to retain all the
properties of the Standard Model Higgs boson.

The MSSM provides a nice illustration of these considerations.  A Higgs
boson of mass 100~GeV (and with properties approximating those of
$\hsm$) is perfectly consistent in the context of the
MSSM.  In this case, the Standard Model breaks down at an energy scale
far below 1000~TeV, due to the existence of supersymmetric
partners whose masses are no heavier than (roughly) 1~TeV.  In
particular, $\hl\simeq\hsm$ in the MSSM as long as $\mha\gsim 2\mz$, as
noted in Section IIIA.

\subsection{Higgs Mass Bounds in Low-energy Supersymmetric Models}

If the minimal supersymmetric extension of the Standard Model (MSSM)
is correct, then we should identify the scale $\Lambda$ at which the
Standard Model breaks down as the scale of low-energy supersymmetry
breaking.  In models of low-energy supersymmetry, $\Lambda$ is
presumed to lie between $m_Z$ and about 1~TeV.
The mass of the light CP-even neutral Higgs boson, $h^0$, in the MSSM
can be calculated to arbitrary accuracy in terms of two parameters of
the Higgs sector, $m_{A^0}$ and $\tan\beta$ \cite{hhg}, and other MSSM
soft-supersymmetry-breaking parameters that affect the Higgs mass
through virtual loops~\cite{hhprl}.  If the
scale of supersymmetry breaking is much larger than $m_Z$, then large
logarithmic terms arise in the perturbation expansion.  These large
logarithms can be resummed using renormalization group (RG) methods.

At tree level, the mass of the lightest CP-even Higgs boson
of the MSSM is bounded: $\mhl\leq\mz|\cos 2\beta|\leq\mz$.
If this prediction were exact, it would imply that the Higgs boson
must be discovered at the LEP-2 collider (running at its projected
maximum
center-of-mass energy of 192~GeV, with an integrated luminosity of
150~${\rm pb}^{-1}$).  Absence of a Higgs boson lighter than $\mz$
would apparently
rule out the MSSM.  When radiative corrections are included, the
light Higgs mass upper bound is increased significantly.  In the
one-loop leading logarithmic approximation \cite{hhprl},
\begin{equation} \label{mhlapprox}
\mhl^2\lsim\mzz\cos^2\beta+{3g^2 m_t^4\over
8\pi^2\mww}\,\ln\left({M^2_{\tilde t}\over \mt^2}\right)\,,
\end{equation}
where $M_{\tilde t}$ is the (approximate) common mass of the
top-squarks.  Observe that the Higgs mass upper bound is very sensitive
to the top mass and depends logarithmically on the
top-squark masses.  Although eq.~(\ref{mhlapprox}) provides a rough
guide to the Higgs mass upper bound, it is not sufficiently precise for
LEP-2 phenomenology, whose Higgs mass reach depends delicately on the
MSSM parameters.  In addition, in order to perform precision Higgs
measurements and make comparisons with theory, a more accurate result
for the Higgs sector masses (and couplings) are required.
The formula for the full one-loop
radiative corrected Higgs mass has been obtained in the literature,
although it
is very complicated since it depends in detail on the virtual
contributions of the MSSM spectrum~\cite{honeloop}.
Moreover, if the supersymmetry breaking scale is larger than a few
hundred GeV, then RG methods are essential for summing up the effects of
large logarithms and obtaining an accurate prediction.

The computation of the RG-improved
one-loop corrections requires numerical integration of a coupled set of
RG equations~\cite{llog}. (The dominant two-loop next-to-leading
logarithmic results are also known~\cite{hempfhoang}.)
Although this program has been
carried out in the literature, the procedure is unwieldy
and not easily amenable to large-scale Monte-Carlo analyses.
Recently, two groups have presented a simple analytic procedure for
accurately approximating $m_{h^0}$.
These methods can be easily implemented, and incorporate both the
leading one-loop and two-loop effects and the RG-improvement.
Also included are the leading effects at one loop of supersymmetric
thresholds (the most important effects of this type are squark mixing
effects in the third generation). Details of the techniques can
be found in Ref.~\cite{hhh} and \cite{carena}, along with other
references to the original
literature.  Here, we simply quote two specific bounds, assuming
$\mt=175$~GeV and $M_{\tilde t}\lsim 1$~TeV:
$\mhl\lsim 112$~GeV if top-squark mixing is negligible, while
$\mhl\lsim 125$~GeV if top-squark mixing is ``maximal''.
%
%
Maximal mixing corresponds to
an off-diagonal squark squared-mass that produces the largest value of
$\mhl$.  This mixing leads to an extremely large splitting of top-squark
mass eigenstates.
Current state of the art calculations can obtain a mass bound
for the light CP-even Higgs boson of the MSSM that
is reliable to within a few GeV.  Of course, the bound one finally
obtains is very sensitive to the top quark mass, and depends crucially
on the upper bound one chooses to place on supersymmetric
particle masses.  In this report, a conservative bound of
$\mhl\lsim 130$~GeV was used as input to the phenomenological
analysis.

The charged Higgs mass is also constrained in the MSSM.  At tree level,
$\mhpm^2=\mw^2+\mha^2$, which implies that charged Higgs bosons cannot
be pair produced at LEP-2.  Radiative corrections modify the tree-level
prediction, but the corrections are typically smaller than the neutral
Higgs mass corrections discussed above.  Although $\mhpm\geq\mw$ is not
a strict bound when one-loop corrections are included, the bound holds
approximately over most of MSSM parameter space (and can be significantly
violated only when $\tanb$ is well below 1, a region of parameter space
that is theoretically disfavored).

The MSSM Higgs mass bounds do not in general apply to
non-minimal supersymmetric extensions of the Standard Model.
If additional Higgs singlet and/or triplet fields are introduced,
then new Higgs self-couplings parameters appear, which are
not significantly constrained by present data.  These parameters can
contribute to the light Higgs masses; \footnote{This should
be contrasted with the MSSM, where all
Higgs self-couplings are related by supersymmetry to
gauge couplings.  This is the origin of the MSSM bound $\mhl\lsim{\cal
O}(\mz)$ discussed above.}
the upper bound on these
contributions depends on an extra assumption beyond the physics
of the TeV scale effective theory.  For example, in the
simplest non-minimal supersymmetric extension of the
Standard Model (NMSSM),
the addition of a Higgs singlet superfield adds a new Higgs
self-coupling parameter, $\lambda$ \cite{nmssm}.
The mass of the lightest neutral Higgs boson can be
raised arbitrarily by increasing the value of $\lambda$ (analogous to
the behavior of the Higgs mass in the Standard Model!).  In this case,
we must generalize the analysis of Section IIIB and introduce
a new scale $\widetilde\Lambda$ beyond which the NMSSM breaks down.  The
upper bound on the Higgs mass then depends on the choice of
$\widetilde\Lambda$.  The standard assumption of theorists who construct
low-energy supersymmetric models is that all couplings stay perturbative
up to the Planck scale.  Choosing $\widetilde\Lambda\simeq\mpl$, one
finds in most cases that $\mhl\lsim 150$~GeV, independent of the
details of the low-energy supersymmetric model
\cite{GKW}.  The NMSSM also permits a tree-level charged Higgs
mass below $\mw$.  However, as in the MSSM, the charged Higgs
mass becomes large and roughly degenerate with $\mha$ in the decoupling
limit where $\mha\gg\mz$.

\section{ESSENTIALS OF HIGGS\\ PHENOMENOLOGY AT FUTURE COLLIDERS}

Higgs hunting at future colliders will consist of three phases.  Phase
one is the initial Higgs boson search in which a Higgs signal is found
and confirmed as evidence for new phenomena not described by
Standard Model background.  Phase two will address the
question: should the signal be identified with Higgs physics?  Finally,
phase three will consist of a detailed probe of the Higgs sector and
precise measurements of Higgs sector observables.  Further details on
the results of this section can be found in Ref.~\cite{gunreport}.

\subsection{Phase 1 -- Demonstrate the Observability \\
of a Higgs Signal}

As we plan for future collider facilities, the machine and detector
characteristics must be developed in such a way that a Higgs signal can
be unambiguously detected above the Standard Model background.
In this discussion, we shall focus mainly
on the Standard Model Higgs boson
($\hsm$) and the Higgs bosons of the MSSM ($\hl$, $\hh$, $\ha$, and
$\hpm$).  At present, taking into account data from LEP-1 and
the most recent LEP-2 data (at $\sqrt{s}=161$ and 172~GeV), one can
exclude a Higgs boson of mass $\mhsm< 70.7$~GeV \cite{ypan}.
The MSSM bounds are a little more complicated, since they
depend primarily on two Higgs sector parameters, but with some
dependence on the MSSM spectrum which affects Higgs masses and couplings
through virtual loop effects.   The current
MSSM Higgs mass bounds exclude the mass
ranges: $\mhl< 62.5$~GeV (independent of the value of $\tan\beta$)
and $\mha< 62.5$~GeV (assuming $\tan\beta> 1$)
\cite{ypan}.  LEP-1 data also excludes charged Higgs masses
with $\mhpm<44$~GeV in a general two-Higgs-doublet model \cite{LEPHIGGS}.
(LEP-2 data does not yet improve this bound.)
This bound is less interesting in the MSSM, where
$\mhpm\gsim\mw$ over most of the MSSM parameter space.
The search for $t\to bH^+$ at the Tevatron can, in principle, extend the
reach of the charged Higgs search.  However, the quoted limits
\cite{CDFHIGGS} apply only in a very narrow region of parameter space.

Consider the Higgs search at future colliders.  The machines we have
examined are summarized in Table~\ref{colliders}.  Most work on
analyzing the
discovery reach of future colliders has focused on the Standard Model
Higgs boson and the Higgs bosons of the MSSM.  In the latter case, some
of the analyses also apply to more general unconstrained versions of
the two Higgs doublet model.   In the decoupling limit, the discovery
limits obtained for $\hsm$ also apply to the lightest CP-even neutral
Higgs boson of a more general non-minimal Higgs sector.

\begin{table}[htb]
\caption{The $\protect \hsm$ discovery reach of future colliders.
A $5\sigma$ signal above background is required for discovery.
Note that Run II at the Tevatron complements the
LEP Higgs search only for an integrated luminosity well beyond one
year at the design luminosity of the Main Injector.
For NLC, both $\protect \sqrt{s}=500$~GeV and 1~TeV cases are shown.
The FMC discovery reach is similar to that of the NLC for the same
center-of-mass energy and integrated luminosity.}
\label{hsmdiscovery}
\renewcommand\arraystretch{1.2}
\setlength{\tabcolsep}{7.5pt}
\begin{center}
\begin{tabular}{lcc}
\hline \hline
                            & Integrated & Discovery \\[-2pt]
\multicolumn{1}{c}{Collider}& Luminosity &  Reach \\
\hline
LEP-2 ($\sqrt{s}=192$~GeV)&    150 pb$^{-1}$ & 95 GeV    \\
Tevatron (M.I.) & 5--10 fb$^{-1}$  & 80--100 GeV \\
TeV-33 & 25--30 fb$^{-1}$ & 120 GeV \\
LHC&     100 fb$^{-1}$ & 800 GeV\\
NLC-500 & 50 fb$^{-1}$ & 350 GeV \\
NLC-1000 & 200 fb$^{-1}$ & 800 GeV \\[2pt]
\hline\hline
\end{tabular}
\end{center}
\end{table}

\noindent\qquad
{\it 1. The Standard Model Higgs Boson}\\

The $\hsm$ discovery reach of future colliders is summarized in
Table~\ref{hsmdiscovery}.  At LEP-2 running at its maximum energy of
$\sqrt{s}=192$~GeV, the discovery
reach of $\mhsm\simeq 95$~GeV can be attained by one detector taking
data for about one year at design luminosity \cite{janot}.  With four
LEP detectors running, the Higgs mass discovery reach can be achieved
sooner (or improve on the significance of any candidate Higgs signal).
Additional luminosity cannot significantly extend the Higgs mass reach
unless the LEP-2 center-of-mass energy were increased.  At
Run II of the Tevatron
one year of data taking at the Main Injector design luminosity is
not sufficient to discover a Standard Model Higgs boson above
background.  However, two detectors running at design luminosity from
three to five years can complement the LEP-2 Higgs search.  In
particular, the associated production of $W\hsm$ with $\hsm\to b\bar b$
may be feasible at the Tevatron, given sufficient integrated
luminosity.  Assuming a
total integrated luminosity of 5 [10] fb$^{-1}$, a Standard Model Higgs
mass discovery reach of 80 [100]~GeV is attainable \cite{tevreport,kky}.
The Tevatron Higgs search technique also applies at higher luminosity.
For example, initial studies indicate that at TeV-33, a
Standard Model Higgs boson with a mass of 120 GeV can be discovered
with an integrated luminosity of 25--30 fb$^{-1}$ \cite{tevreport,kky}.
The significance of the Higgs signal could be enhanced by the detection
of the associated production of $Z\hsm$, $\hsm\to b\bar b$ \cite{yao}.
Implicit in these studies is the assumption that the Standard
Model contributions are sufficiently well understood that the Higgs
signal can be detected as a small excess above background.

The LHC is required if one wants to extend the Higgs mass discovery
reach significantly beyond ${\cal O}(m_Z)$.  Note that according to the
discussion of Section IIIB, it only makes sense to consider Standard
Model Higgs bosons with mass below 800~GeV.\footnote{It is possible to
imagine theories of electroweak symmetry breaking which produce scalar
states heavier than 800~GeV.  However, any such scalar is presumably
either strongly coupled, and/or composite on the scale of 1~TeV.
The consideration of such scalars
lies outside the scope of our working group.}
Therefore,
Table~\ref{hsmdiscovery} implies that the LHC can provide complete
coverage of the (weakly-coupled) Standard Model Higgs mass region,
assuming that it achieves its design luminosity \cite{gunreport,atlas,cms}.
For $\mhsm\gsim 2\mz$, the ``gold-plated mode'' $\hsm\to
ZZ\to\ell^+\ell^-\ell^+\ell^-$ provides a nearly background free
signature for Higgs boson production until the production rate becomes
too small near the upper end of the weakly-coupled Higgs mass regime.
In this case, other signatures ({\it e.g.}, $\hsm\to ZZ\to
\ell^+\ell^-\nu\bar\nu$ and $\hsm\to W^+W^-\to \ell\nu+{\rm jets}$)
provide additional signatures for Higgs discovery.

The most troublesome Higgs mass range for hadron colliders
is the so-called
``intermediate Higgs mass regime'', which roughly corresponds to
$\mz\lsim\mhsm\lsim 2\mz$.  For 130~GeV$\lsim\mhsm\lsim 2\mz$, one can
still make use of the gold plated mode at the LHC,
$\hsm\to ZZ^{*}\to\ell^+\ell^-\ell^+\ell^-$ (where $Z^{*}$ is virtual).
Standard
Model backgrounds begin to be problematical when the branching ratio
${\rm BR}(\hsm\to ZZ^{*})$ becomes too small.  This occurs for
$2\mw\lsim\mhsm\lsim 2\mz$ where ${\rm BR}(\hsm\to W^+W^-)$ is
by far the dominant Higgs decay channel, and for
$\mhsm\lsim 140$~GeV where the the virtuality of $Z^{*}$ begins to
significantly reduce the $\hsm\to ZZ^{*}$ decay rate.
A complementary channel $\hsm\to WW^{(\ast)}\to \ell^+\nu\ell^-\bar\nu$
provides a viable Higgs signature for 155~GeV$\lsim\mhsm\lsim 2\mz$
\cite{Dreiner}, and closes a potential hole near the upper end of the
intermediate Higgs mass range.  For $\mhsm\lsim
130$~GeV, the dominant decay channel $\hsm\to b\bar b$ has very large
Standard Model two-jet backgrounds. Thus, in this regime, it is
necessary to consider rarer production and decay modes with more
distinguishing characteristics.  Among the signatures studied in the
literature are:
\begin{list}{(\roman{mynum})}{\usecounter{mynum} \itemsep 0in
\labelwidth .5in \leftmargin .7in}
\item $gg\to\hsm\to\gamma\gamma$,
\item $q\bar q\to V^{*} \to V\hsm$, \qquad ($V=W$ or $Z$),
\item $gg\to t\bar t\hsm$,
\item $gg\to b\bar b\hsm$, and
\item $gg\to\hsm\to\tau^+\tau^-$.
\end{list}
The LHC detectors are being optimized
in order to be able to discover an intermediate mass Higgs boson via its
rare $\gamma\gamma$ decay mode (with a branching ratio of about
$10^{-3}$).  The other signatures could be used to
provide consistency checks for the Higgs discovery as well as provide
additional evidence for the expected Higgs-like properties of the Higgs
boson candidate.  A successful intermediate mass Higgs search
via the $\gamma\gamma$ decay mode
at the LHC will require maximal luminosity and
a very fine electromagnetic calorimeter resolution
(at about the 1\% level).

In contrast to the Tevatron and LHC Higgs searches, the
Standard Model Higgs search at the NLC in the intermediate mass regime
is straightforward, due to the simplicity of the Higgs
signals, and the relative ease in controlling the Standard Model
backgrounds. Higgs production is detected at the NLC via two main
signatures.  The first involves the extension of the LEP-2 search for
\begin{equation}
e^+e^-\to Z\hsm
\label{BJ}
\end{equation}
to higher energies.
In addition, a second process can also be significant:
the (virtual) $W^+W^-$ fusion process\footnote{The corresponding $ZZ$
fusion process, $e^+e^-\to e^+e^-Z^*Z^*\to e^+e^-\hsm$ is suppressed
by about a factor of ten relative to the $W^+W^-$ fusion process.
Nevertheless, at large $\sqrt{s}/\mhsm$, the $ZZ\to\hsm$ fusion rate
compares favorably to that of $e^+e^-\to Z\hsm$.  As a result, the $ZZ$
fusion process can be used in some cases to study Higgs properties.}
\begin{equation}
e^+e^- \to\nu\bar\nu W^* W^* \to\nu\bar\nu\hsm.
\label{FUSSION}
\end{equation}
The fusion cross-section grows logarithmically with the center-of-mass
energy and becomes the dominant Higgs production
process at large $\sqrt{s}/\mhsm$. For example, at
$\sqrt{s}=500$~GeV, complete coverage
of the intermediate Higgs mass regime below $\mhsm\lsim 2\mz$ requires
only 5~fb$^{-1}$ of data.  The only limitation of the NLC in the Higgs
search is the center-of-mass energy of the machine which
determines the upper limit of the Higgs boson discovery reach.
One would need $\sqrt{s}\simeq 1$~TeV
to fully cover the weakly-coupled Standard Model Higgs
mass range \cite{finland,nlcreport,desyreport}.

The techniques for the Standard Model Higgs boson {\it discovery} at a
$\mu^+\mu^-$
collider are, in principle, identical to those employed at the
NLC \cite{BBGH,mumureport}.  However,
one must demonstrate that the extra background resulting from an
environment of decaying muons can be tamed.  It is believed that
sufficient background rejection can be achieved \cite{miller}; thus
the FMC has the same discovery reach as the NLC at the same
center-of-mass energy and luminosity. \\


\noindent\qquad
{\it 2. Higgs Bosons of the MSSM}\\

Next, we turn to the discovery potential at future colliders for the
Higgs bosons of the MSSM.  If $\mha\gg\mz$, then the decoupling limit
applies, and the couplings of $\hl$ to Standard Model particles are
identical to those of $\hsm$.  Thus, unless $\hl$ decays appreciably to
light supersymmetric particles, the discussion given above for $\hsm$
apply without change to $\hl$.  In general, one can consider two types
of MSSM Higgs searches at future colliders.  First, one can map out the
region of MSSM parameter space where at least one MSSM Higgs
boson can be discovered in a future collider Higgs search.  If no Higgs
state is discovered, then the corresponding region of MSSM parameter
space would be excluded.  (In some cases, the
absence of a Higgs discovery would be strong enough to completely rule
out the MSSM!)
Note that in this approach, one may simply discover one
Higgs state---the light CP-even neutral $\hl$---with properties
resembling that of $\hsm$, which would be consistent
with MSSM expectations, but would provide no direct proof that
low-energy
supersymmetry underlies the Higgs sector dynamics.
Second, one can examine the discovery potential for
specific states of the non-minimal Higgs sector.  As emphasized in
Section IIIA, in the decoupling limit, the non-minimal Higgs states are
heavy (compared to the $Z$), nearly degenerate in mass, and
weakly-coupled.  Discovery of these states at future colliders is far
from being assured.

\setlength{\tabcolsep}{5pt}
\begin{table}[htb]
\caption{MSSM Higgs boson discovery potential}
\label{mssmdiscpotential}
\vskip-1pc
\renewcommand\arraystretch{1.2}
\setlength{\tabcolsep}{4pt}
\begin{center}
\begin{tabular}{lp{7cm}}
\hline\hline
\multicolumn{1}{c}{Collider}&
\multicolumn{1}{c}{Comments}\\[2pt]
\hline
LEP-2& Significant but not complete coverage, via\\
     & \qquad $\epem\to H^+H^-$ \hfill\break
       \null\qquad $\epem\to Zh^0$ \hfill\break
       \null\qquad $\epem\to h^0A^0$ \\[3pt]
TeV-33& Limited coverage, complements the
          LEP-2 search \\[3pt]
LHC&  (Nearly) complete coverage for the discovery of at
      least one Higgs boson of the MSSM.
      Main challenge:  the intermediate Higgs mass region
      [$m_Z \lsim m_{h^0} \lsim 2m_Z$] which requires different
      search strategies depending on the value of $m_{h^0}$.\hfill\break
      Some sensitivity to heavier non-minimal Higgs states.\\[3pt]
NLC&  Complete coverage for the discovery of at least \\[-3pt]
and&  one Higgs boson of the MSSM.  Sensitivity to \\ [-3pt]
FMC&      to heavier non-minimal
      states depends on $\sqrt s$:  \hfil\break
     $\sqrt s\gsim 2m_A$ \quad
            for discovery of $H^\pm,H^0,A^0$ via  \hfil\break
     \hphantom{%
     $\sqrt s\sim 2m_A \quad\;$} \qquad associated production. \\
&    $\sqrt s\sim  m_A \9\quad$
            for $\mu^+\mu^- \to H^0,A^0$ $s$-channel \hfil\break
     \hphantom{%
     $\sqrt s\sim m_A \9\quad$} \qquad resonance production.   \\
\hline \hline
\end{tabular}
\end{center}
\end{table}

We summarize the MSSM Higgs boson discovery potential at future colliders
in Table~\ref{mssmdiscpotential}.\footnote{We have not considered the
possibility of Higgs decay channels involving supersymmetric particles.
This is probably not an issue for the lightest CP-even scalar, $\hl$.
Recall that in the MSSM, $\mhl\lsim 130$~GeV, and consider the likely
constraints on supersymmetric particle masses in the absence of observed
supersymmetric particle production at LEP-2.  It is then very unlikely
that there would be any open supersymmetric channels in $\hl$ decays.
For the heavier Higgs states ($\hh$, $\ha$ and $\hpm$), supersymmetric
decay modes can be significant and provide new
signatures for Higgs production and decay. This possibility merits
further study.}
Consider first the discovery limits for $\hl$ of the MSSM at
future collider facilities.  As described in Section IIIC, the tree-level
MSSM predicts that $\mhl\leq\mz$.  Suppose that this predicted bound
were unmodified (or reduced) after taking radiative corrections into
account.  Then the non-observation of $\hl$ at LEP-2 (which will
eventually be sensitive to the mass range $\mhl\lsim
95$~GeV) would rule out the MSSM.  However, for some choices of MSSM
parameters, the radiative corrections
significantly {\it increase} the tree-level bound \cite{hhprl}.
Consequently, the Higgs searches
at LEP-2 (and the Tevatron) cannot completely rule out the MSSM.

On the other hand, considering that the radiatively
corrected bound is $\mhl\lsim 130$~GeV, it would appear that the LHC has
access to the full MSSM Higgs sector parameter space.  After all, we
argued above that the LHC will be able to completely cover the
intermediate Standard Model Higgs mass regime.  However, when
$\mha\sim{\cal O}(\mz)$,
the decoupling limit does not apply, and the properties of $\hl$ deviate
from those of $\hsm$.  Thus, an independent analysis is required
to ascertain the discovery potential of the LHC search for MSSM Higgs
bosons.  In particular, the LHC detector collaborations must demonstrate
the feasibility of $\hl$ discovery in the mass range $\mz\lsim\mhl\lsim
130$~GeV.  This is precisely the most difficult region for the LHC Higgs
search.  At this time, one can argue that the LHC coverage of the MSSM
Higgs sector parameter space is nearly complete, although the
search strategies sometimes depend on the observation of small signals (above
significant Standard Model backgrounds) in more than one channel.
Moreover, the present estimates of the statistical significance of the
Higgs signal rely on theoretical determinations of both signal and
background rates as well as simulations of detector performance.  Thus,
if no Higgs signal is confirmed by the LHC, it might still be difficult
to definitively rule out the MSSM.

The NLC (and FMC) provide complete
coverage of the MSSM Higgs sector parameter space once the
center-of-mass energy is above 300~GeV.  In contrast to the LHC
Higgs search, the intermediate Higgs mass regime presents no particular
difficulty for the high energy lepton colliders.  The associated
production
\begin{equation}
e^+e^-\to\hl\ha
\label{ha}
\end{equation}
provides an addition discovery channel for $\mha\lsim\sqrt{s}/2$.
If no Higgs signal is seen, then the
lepton colliders can unambiguously rule out the MSSM. \\

\noindent\qquad
{\it 3. Higgs Bosons in non-minimal extensions of the MSSM}\\

If no Higgs state is discovered at the LHC and NLC, then the MSSM would
cease to be a viable candidate for a theory of electroweak physics.
However, the MSSM is just one model of low-energy supersymmetry.  Thus,
it is important to consider non-minimal extensions of the MSSM to see
whether the low-energy supersymmetric approach could be ruled out in
general.  Consider the Higgs search at the NLC in the context of
a completely general two-Higgs doublet model.
Suppose that the non-minimal Higgs states are heavy so that only
$\hl$ is accessible at the NLC.  The relevant $\hl$ production processes
are listed in Eqs.~(\ref{BJ}) and (\ref{FUSSION}).
Note that in both cases, the production cross-sections are governed by
the strength of the $\hl$ coupling to vector boson pairs.  But, in
models with Higgs doublets and singlets (but with no higher Higgs
multiplets), these couplings must satisfy a sum rule \cite{ghw}:
\begin{equation}\label{sumrule}
\sum_i\,g^2_{VV\hl_i}=g^2_{VV\hsm}\,,
\end{equation}
where $V=W^\pm$ or $Z$.  As an example,
the $5\sigma$ discovery of the Standard Model
Higgs boson with $\mhsm=150$~GeV at the NLC running at
$\sqrt{s}=500$~GeV requires only about 2~fb$^{-1}$ of data
(see, {\it e.g.}, Fig. 2.18 of Ref.~\cite{nlcreport}), corresponding
to about 100 Higgs boson events before cuts.
Equivalently, the NLC running at $\sqrt{s}=500$~GeV with an integrated
luminosity of 50~fb$^{-1}$ permits the 5$\sigma$ discovery of
a neutral CP-even Higgs boson with 4\% of the Standard Model cross section,
which corresponds to $g_{VV\hl}\gsim 0.2 g_{VV\hsm}$.
Of course, if the $VV\hl$ coupling
were smaller than this, no Higgs state would be discovered in this
experiment.  However, by raising the center-of-mass energy of the NLC, one
must eventually find evidence for at least one of the heavier neutral
Higgs states, by virtue of the sum rule [eq.~(\ref{sumrule})] quoted above.


The situation where the bulk of the $VV\hl_i$ couplings are carried by
the heavier Higgs states cannot arise in the MSSM for two reasons.
First, the MSSM Higgs mass bound implies that $\mhl\lsim 130$~GeV, and
second, $g_{VV\hl}\lsim 0.2g_{VV\hsm}$ is possible only if $\mha\lsim
{\cal O}(\mz)$, in which case, the Higgs boson would be discovered
via $\hl\ha$ production [eq.~(\ref{ha})].
In non-minimal extensions of the MSSM Higgs
sector, these two objections must be reconsidered.

We reviewed
the case of the NMSSM in which one complex Higgs singlet field is added.
This model introduces a new independent Higgs self-coupling which {\it a
priori} can take on any value.  However, if one imposes the requirement
of perturbativity of couplings at all scales below the Planck scale (a
requirement motivated by the unification of strong and electroweak
couplings near the Planck scale), then one finds that the lightest Higgs
boson must satisfy $\mhl\lsim 150$~GeV.  Still, the lightest CP-even
Higgs scalar may be very weakly coupled to quarks, leptons and gauge
bosons if it is primarily composed of the singlet component.  Thus, a
detailed analysis is required to see whether the Higgs search at the NLC
is sensitive to all regions of the NMSSM Higgs sector parameter space.
The analysis of Ref.~\cite{KOT} demonstrated that even for
$\sqrt{s}=300$~GeV, the NLC search would easily detect at least one
Higgs state of the NMSSM.  Specifically, the minimum Higgs production
cross-section in the NMSSM at $\sqrt{s}=300$~GeV [500~GeV] was found to
be 42~fb [17~fb].  Such Higgs production rates are easily detected above
background, assuming the NLC luminosity given in Table~\ref{colliders}.

A similar question can be posed in the case of the LHC Higgs search.
As discussed earlier,
the LHC search will provide nearly complete coverage of the MSSM
Higgs sector parameter space.
Nevertheless, the LHC search is
operating ``at the edge'' of its capabilities.  By relaxing some of the
MSSM constraints to Higgs sector parameters, we expect some holes to
develop in region of supersymmetric
parameter space accessible to the LHC Higgs search.
Ref.~\cite{GHM} examined this question in detail for the case of
the NMSSM, and concluded that although the region of inaccessibility is
not large, it is possible to find regions of NMSSM Higgs parameter space
in which no Higgs boson state could be discovered at the LHC.
This analysis does suggest the possibility that future improvements in
search strategies and detector capabilities
(for example, improved $b$-tagging) may be able to significanly narrow the
region of inaccessibility in the Higgs sector parameter space.
Clearly, the supersymmetric Higgs search
remains a formidable challenge for future experimentation at LHC.

The above considerations can also be applied to more general non-minimal
extensions of the MSSM.  Although there is no completely general
analysis yet available, under
most reasonable model assumptions, the non-observation
of a Higgs boson in the intermediate-Higgs-mass regime at the NLC would
rule out the low-energy supersymmetric model.  Whether this
``no-go'' theorem can be circumvented by some more exotic approach to
low-energy supersymmetry remains to be seen. \\

\noindent\qquad
{\it 4. Observing More Than One Higgs Boson}\\

If only one
Higgs boson is discovered, it may closely resemble the
$\hsm$.  In this case, one must address the detectability of the
non-minimal Higgs states ($\hh, \ha, \hpm, \cdots$) at future
colliders.  As emphasized above, all future colliders can provide only
incomplete probes of the non-minimal Higgs sector parameter space.
Naively, one would expect the masses of all Higgs sector states to be of
order the electroweak scale.  However, somewhat heavier non-minimal
Higgs states often arise in model building.  As an example, in
low-energy supersymmetric models, the mass scale of the non-minimal
Higgs states is controlled by a soft-supersymmetry-breaking parameter
which could be as large as 1~TeV.  Such heavy states would still be
weakly-coupled and difficult to observe at any of the colliders we have
examined.

The exploration of the non-minimal Higgs sector parameter space at
future colliders could be especially challenging.  Detection
of heavy non-minimal Higgs states at the LHC is difficult
due to the very low signal-to-background ratio of the
corresponding Higgs boson signals. In
particular, heavy Higgs states couple very weakly to gauge bosons, and
would have to be detected via their heavy fermion decays.  (At large
$\tan\beta$, where the Higgs couplings to down-type fermions is
enhanced relative to the Standard Model, it may be possible to observe a
heavy neutral Higgs boson via its decay to $\tau^+\tau^-$.)
At the NLC, the main obstacle for the discovery of non-minimal Higgs
states is the limit of the center-of-mass
energy.  For reasons connected to the nature of the
decoupling limit, the heavy Higgs states of the MSSM can be produced in
sufficient number and detected only if $\sqrt{s}\gsim 2\mha$
\cite{DECP}. The
discovery reach could in principle be somewhat extended by employing the
$\gamma\gamma$ collider mode of the NLC.
In this mode of operation, the search for $\gamma\gamma\to\ha$ and
$\gamma\gamma\to\hh$ can
extend the non-minimal Higgs mass discovery
reach of the NLC \cite{gunhab}.

Finally,
the FMC can produce the neutral Higgs states singly via $s$-channel
$\mu^+\mu^-$ annihilation, and would permit the discovery of the
heavy neutral Higgs states up to $\sqrt{s}=\mha$ \cite{BBGH}.
The viability of this discovery mode depends on the parameters of the
Higgs sector.  In the MSSM, the cross-section for
$\mu^+\mu^-\to \hh,\ha$
is enhanced for values of $\tan\beta$ above 1.  For $\mhh, \mha\gg\mz$,
$\hh$ and $\ha$ are approximately degenerate in mass.
Given sufficient luminosity, one can detect $\hh$ and
$\ha$ (if kinematically accessible) by scanning in $\sqrt{s}$, assuming
that $\tan\beta$ is larger than a critical value (which depends on the
total luminosity and the Higgs mass).  Detection is accomplished via a
resonant peak in the Higgs decay to $b\bar b$ (and $t\bar t$ if
allowed).  Further details can be found in Ref.~\cite{gunreport}.

\subsection{Phase 2 -- After Discovery: Is It a Higgs Boson?}

Suppose that the first candidate Higgs signal is detected.
What must one do to prove that the produced state is a Higgs boson?
We assume that after the initial discovery is made, further collider
running confirms the signal and establishes a useful statistical sample
of events.  The first step is to ascertain whether
the observed state resembles the Standard Model Higgs boson and/or if
it is associated with a non-minimal Higgs
sector.  If $\hl\simeq\hsm$, then one must demonstrate that the
discovered state has
\begin{list}{(\roman{mynum})}{\usecounter{mynum} \itemsep 0in
\labelwidth .5in \leftmargin .7in}
\item zero electric and color charge,
\item spin zero,
\item CP-even quantum number,
\item electroweak strength couplings, and
\item couplings proportional to the mass of the state to which it
couples.
\end{list}
Eventually, one would like to make detailed
measurements and
verify that the Higgs candidate matches all the properties expected of
$\hsm$ to within some precision (small deviations from the $\hsm$
properties
will be addressed in the next section).  If the properties of the
discovered state are Higgs-like, but differ in detail from those of
$\hsm$, then it is likely that other non-minimal Higgs states are light
and may have been produced in the same experiment.  Finding evidence for
these states will be crucial in verifying the Higgs interpretation of
the data.

At an $e^+e^-$ collider (LEP-2 and the NLC),
many of the Higgs boson properties can be directly measured due to
low backgrounds and simple event structures.\footnote{In principle,
the remarks that follow also apply to the FMC.  However, it has not yet
been demonstrated that the severe backgrounds arising from the
constantly decaying muons can be overcome to make precision
measurements.}
One can directly
measure the spin and CP-quantum numbers of the Higgs candidate through
the angular distributions of production and decay.  Specific Higgs decay
modes can be separated and individually studied.  Accurate measurements
of $\sigma(\hl){\rm BR}(\hl\to X)$ can be made for a number of final
states, including $X=b\bar b$ and $\tau^+\tau^-$.  In this workshop, a
breakthrough was reported
demonstrating that the detection of $\hl\to c\bar c$ is possible
with appreciable efficiency and low mis-identification \cite{bctags}.
Thus, at the lepton colliders, $\hl\simeq\hsm$ can be
confirmed with some precision.

\begin{table}[htb]
\caption{Detectability of the $\hsm$ at future hadron colliders
as a function of the $\hsm$ mass range.
For the Tevatron Higgs search (Run~II of the Main Injector and the
TeV-33 upgrade), the required integrated luminosity in
units of fb$^{-1}$ is indicated in braces.  For comparison,
the LEP-2 discovery range (via
$e^+e^-\to Z\hsm$) is indicated.   For the Tevatron and LHC searches,
the Higgs decay modes involved in the primary Higgs discovery
signals are shown in parentheses; further details are given in
Table~\ref{tab:hsmsignatures}.}
\label{massranges}
\vskip-6pt
\renewcommand\arraystretch{1.2}
\setlength{\tabcolsep}{9pt}
\begin{center}
\begin{tabular}{ll}
\hline \hline
\multicolumn{1}{c}{Mass Range} &
          \multicolumn{1}{c}{Observability at Future Colliders} \\
\hline
60--80 GeV&   LEP-2, Tevatron\{5\}($b\bar b$) \\
80--100 GeV&  LEP-2, Tevatron\{10\}($b\bar b$), and LHC($\gamma\gamma$)
\\
100--120 GeV& Tevatron\{25--30\}($b\bar b$) and LHC($\gamma\gamma$) \\
120--130 GeV& LHC ($\gamma\gamma$) \\
130--155 GeV& LHC ($ZZ^{*}$) \\
155--180 GeV& LHC ($ZZ^{*}, W^+W^-$) \\
$\gsim 180$ GeV & LHC ($ZZ\to\ell^+\ell^-\ell^+\ell^-,
\ell^+\ell^- \nu \bar \nu$) \\
\hline  \hline
\end{tabular}
\end{center}
\vspace*{-6pt}
\end{table}

The initial Higgs discovery is most likely to occur at either LEP-2 or
LHC.  Thus, it is important to examine whether it is possible
to verify the Higgs interpretation of a Higgs signal discovered at the
approved future facilities.  A strategy for accomplishing
this goal was developed by our working
group. We considered what was achievable on the basis of the Higgs
searches at
LEP-2, Run II at the Tevatron
(with some consideration of a possible TeV-33 upgrade\footnote{The Higgs
discovery reach at the Tevatron depends critically on the total
integrated luminosity processed and analyzed by the CDF and D0
detectors.  The Tevatron at the Main Injector design
luminosity must run with one detector for five years
to attain a Higgs discovery reach up to 100 GeV.  To extend the Higgs
reach further before the start of LHC requires TeV-33.})
and the LHC.   We elucidated all the observables where a Higgs
signal could be detected.  We considered separately seven specific mass
intervals in the range 60~GeV $\lsim\mhl\lsim 800$~GeV, listed in
Table~\ref{massranges}.  We then considered in detail a variety of
possible Higgs signatures at each collider
(see Table~\ref{tab:hsmsignatures}) and evaluated the potential
of each channel for supporting the Higgs interpretation of the signal.
Taken one by one, each channel provides limited information.
However, taken together, such an analysis might provide a strong
confirmation of the Higgs-like properties of the observed state as well
as providing a phenomenological profile that could be compared to the
predicted properties of the Standard Model Higgs boson.  Finally, we
considered the limitations of the data from the
Higgs searches at the hadron colliders, and examined the possible
improvements in the determination of the Higgs properties with new data
from the NLC and/or the FMC.
A list of the primary Higgs signals at future colliders considered above
is given in Table~\ref{tab:hsmsignatures}.

\begin{table}[htb]
\caption{Primary $\hsm$ signatures at future colliders and the
corresponding Higgs mass range over which detection of a
statistically significant signal is possible.  Other Higgs signatures
not included in this table are discussed in Ref.~\cite{gunreport}.}
\label{tab:hsmsignatures}
\renewcommand\arraystretch{1.5}
\setlength{\tabcolsep}{5.5pt}
\begin{center}
\begin{tabular}{lll}
\hline\hline
\multicolumn{1}{c}{Collider}&
\multicolumn{1}{c}{Signature}&
\multicolumn{1}{c}{Mass Range}\\[2pt]
\hline
LEP-2&    $e^+e^-\to Z\hsm$&   $\lsim 95$ GeV \\
TeV-33$^a$& $ W^*\to W\hsm\to \ell\nu b\bar b$& 60--120 GeV \\
  & $Z^*\to Z\hsm\to \left\{\matrix{\lplm b\bar b \cr\noalign{\vskip3pt}
                          \nu\bar\nu\ b\bar b\quad}\right.$
                                            &   \\[3pt]
LHC&  $W^*\to W\hsm\to \ell\nu b\bar b$& 80--100 GeV \\
   & $\hsm + X\to \gamma\gamma+X$& 90--140 GeV \\
   & $\hsm\to ZZ^* \to\lplm\lplm$& 130--180 GeV \\
   & $\hsm\to WW^*\to \ell^+\nu\ell^-\bar\nu$& 155--180 GeV \\
   & $\hsm\to ZZ\to \lplm\lplm$& 180--700 GeV \\
   & $\hsm\to ZZ \to \nu\bar\nu\lplm$& 600--800 GeV \\
   & $\hsm\to W^+W^- \to \ell\nu+$jets& 600--800 GeV$^b$\\[3pt]
$\matrix{\rm NLC \cr \hphantom{NLC}\cr \hphantom{NLC} }$&
$\left.\matrix{\!\epem \to Z\hsm\quad\cr\noalign{\vskip3pt}
     \! \epem\to \nu\bar\nu\hsm\quad \cr\noalign{\vskip3pt}
     \! \epem\to \epem\hsm} \right\}$& $\lsim 0.7 \sqrt s$ \\[1.6pc]
$\matrix{\rm FMC\cr \hphantom{FMC}\cr \hphantom{FMC} }$&
$ \left.\matrix{\!\mpmm \to Z\hsm \quad\;\cr \noalign{\vskip3pt}
         \! \mpmm\to \nu\bar\nu\hsm \quad\;\cr  \noalign{\vskip3pt}
         \! \mpmm\to \mpmm\hsm } \right\}$& $\lsim 0.7 \sqrt s$ \\[2pt]
   & $\mpmm\to\hsm$                  &up to $\sqrt s<2\mw$ \\ [3pt]
\hline \hline
\end{tabular}
\end{center}
\vskip-3pt
\noindent $^a$ The TeV-33 Higgs signatures
listed above are also relevant for lower luminosity Tevatron searches
over a more restricted range of Higgs masses, as specified in
Table~\ref{massranges}.

\noindent $^b$ Ref.~\cite{luc} argues that the $\ell\nu$+2 jets
signal can be detected for Higgs masses up to 1~TeV (although such
large Higgs masses lie beyond the scope of this working group).
\end{table}

In order to determine the true identity of the Higgs candidate,
it is very important to be able to detect the Higgs
signal in at least two different channels.
As one can discern from Table~\ref{massranges}, the most problematical
mass range is 100~GeV$\lsim\mhsm\lsim 130$~GeV.  Higgs bosons in this
mass range are not accessible to LEP-2 or Run II of the Tevatron.  At
the LHC, the most viable signatures in this mass range involve the
production of $\hsm$
followed by $\hsm\to\gamma\gamma$.  However, the Higgs can be produced
via a number of different possible mechanisms:
\begin{list}{(\roman{mynum})}{\usecounter{mynum} \itemsep 0in
\labelwidth .5in \leftmargin .7in}
\item $gg\to\hsm$,
\item $q\bar q\to q\bar q\hsm \;$
via $t$-channel $W^+W^-$ fusion,
\item $q\bar q\to V\hsm\; $ via
$s$-channel $V$-exchange, and
\item  $gg\to t\bar t\hsm$.
\end{list}
The $gg\to\hsm$ mechanism dominates, and it will be an experimental
challenge to separate out the other production mechanisms.
It may be possible to separate $gg\to\hsm$ and $W^+W^-\to\hsm$ events
using a forward jet tag which would select out the $W^+W^-$ fusion
events.  It may also be possible to distinguish $V\hsm$
($V=W^\pm$ or $Z$) and $t\bar
t\hsm$ events based on their event topologies. If these other
production mechanisms can be identified,
then it would be possible to extract information about
relative couplings of the Higgs candidate to $VV$ and $t\bar t$.
Otherwise, one will be
forced to rely on matching $\sigma(\hsm){\rm BR}(\hsm\to\gamma\gamma)$
to Standard Model expectations in order to confirm the Higgs
interpretation of $\hsm$.

In some circumstances, it might be possible to observe the decays
$\hsm\to b\bar b$ or $\hsm\to\tau^+\tau^-$ (after a formidable
background subtraction), or identify the Higgs boson produced via $gg\to
b\bar b\hsm$.  One could then extract the relative coupling strengths
of $\hsm$ to $b\bar b$ and/or $\tau^+\tau^-$ final states.  These could
be compared with the corresponding $VV$ and $t\bar t$ couplings (see
above), and confirm that the
Higgs candidate couples to particles with coupling strengths proportional
to the particle masses.

The quantum numbers of the Higgs candidate may be difficult to measure
directly at a hadron collider.
However, note that if $\hsm\to\gamma\gamma$ is seen, then the $\hsm$
cannot be spin-1 (by Yang's theorem).  This does not prove that $\hsm$
is spin-zero, although it would clearly be the most likely possibility.
If the coupling $\hsm VV$ is seen at a tree-level
strength, then this would confirm the presence of a CP-even component.
Unfortunately, any CP-odd component of the state couples to $VV$ at the
loop level, so one would not be able to rule out {\it a priori} a
significant CP-odd component for $\hsm$.

To summarize, Phase 2 consists of determining whether the Higgs
candidate (discovered in Phase 1) can be identified as a Higgs boson.
In some Higgs mass ranges, LEP-2, the Tevatron, and/or the LHC will
discover the Higgs boson and make a convincing case for the ``expected''
Higgs-like properties.  Ratios of Higgs couplings to different final
states may be measured to roughly 20--30\%.  The NLC (and perhaps the
FMC) can make more precise measurements of branching ratios and can
directly check the spin and CP-quantum number of the Higgs candidate.
The lepton machines (with $\sqrt{s}\gsim 300$~GeV) can easily handle the
intermediate Higgs mass regime and can provide valuable information in
some mass regions that present difficulties to hadron colliders.

\subsection{Phase 3 -- Precision Measurements of Higgs Properties}

Let us suppose that the Higgs candidate (with a mass no larger than a
few times the $Z$ mass)  has been confirmed to have the
properties expected of the $\hsm$ (to within the experimental error).
One would then be fairly confident that the
dynamics that is responsible for electroweak symmetry breaking is
weakly-coupled.
Unfortunately, the details of the underlying physics responsible for
electroweak symmetry breaking would still be missing.
As discussed in Section III, it is not difficult to construct
models of the
scalar dynamics that produce a light scalar state with the properties
of the $\hsm$.  To distinguish among such models,
additional properties of the scalar sector must be
uncovered. It is the non-minimal Higgs states that encode the structure
of the electroweak symmetry breaking dynamics.
In order to provide experimental proof of the existence of a non-minimal
Higgs sector, one must either demonstrate that the properties of $\hl$
differ (even if by a small amount) from those of $\hsm$, or one must
directly produce and detect the heavier Higgs states ($\hh, \ha,
\hpm, \cdots$).  In general, precision measurements of both light
and heavy Higgs properties
are essential for distinguishing among models of electroweak symmetry
breaking dynamics.

A precision measurement of the lightest Higgs mass could be useful.
As noted in Section II, the Higgs mass
measurement can provide a non-trivial check of the precision electroweak
fits in the context of the Standard Model.  This analysis would be
sensitive to one-loop (and some two-loop) virtual effects.  Any
significant discrepancy would indicate
the need for new physics beyond the Standard Model.  In this context, a
Higgs mass measurement with a relative error of about 20\% is all that
is required.  In the MSSM, the light Higgs mass measurement provides an
additional
opportunity.  In Section IIIC, it was noted that the light Higgs mass in
the MSSM at tree-level is a calculable function that depends on two Higgs
sector parameters.  When one-loop effects are included, the Higgs mass
becomes dependent upon additional MSSM parameters (the most important
of which are the top-squark masses and mixing parameters).  Since the
radiative corrections to the Higgs mass can be significant, a precision
measurement of the Higgs mass could provide a very sensitive test of
the low-energy supersymmetric model.  Theoretical calculations
yield a prediction for the light CP-even neutral Higgs mass
(which depends on the MSSM parameters), with an
accuracy of about 2 to 3 GeV \cite{hhh,carena}.
The anticipated experimental accuracy of the light Higgs mass
measurement depends on the Higgs mass range and the collider.
Table~\ref{mhprecision} lists the estimated
errors in the measurement of the Standard Model Higgs mass,
$\Delta\mhsm$, at future colliders for $\mw\lsim\mhsm\lsim 2\mw$.
Note that the numbers quoted in Table~\ref{mhprecision} are considerably
{\it smaller} than the theoretical uncertainties quoted above.

\begin{table}[h]
\caption{Anticipated experimental error in the measured value of the
Standard Model Higgs mass, $\protect\Delta\mhsm$, in units of MeV, for
various ranges of $\mhsm$.  The notation ``?'' indicates that a
reliable simulation or estimate is not yet available, while ``--'' means
that the corresponding Higgs mass range is not accessible.
The assumptions underlying the various collider runs listed below are
specified in the text.  See Ref.~\protect\cite{gunreport} for further
details.}
\vspace*{-1pc}
\renewcommand\arraystretch{1.2}
\setlength{\tabcolsep}{6.25pt}
\begin{center}
\begin{tabular}{lcccc}
\hline
\hline   \noalign{\vskip3pt}
       & \multicolumn{4}{c}{$\mhsm$ range (GeV)}\\[2pt]\cline{2-5}
\multicolumn{1}{c}{Collider}& 80 & $\protect\mz$ & 100--120 & 120--150\\[3pt]
\hline \noalign{\vskip3pt}
LEP-2 \protect\cite{patrick}  & $250$ & $400$ & $-$ & $-$ \\
TeV-33  & 960 & $?$ & 1500--2700 & $-$ \\
LHC     & 90 & 90 & 95--105 & 105--90 \\
NLC (500) & 370 & 264 & 200--120 & 120--70 \\
NLC ($\protect\sqrt{s_{Zh}}$)
          & 3.6 & 3.8 & 4.1--4.8 & 4.8--6.1 \\
NLC (threshold) & 40 & 70 & 55--65 & 65--100 \\
FMC (scan) & 0.025 & 0.35 & 0.1--0.06 & 0.06--0.49 \\
\hline
\hline
\end{tabular}
\end{center}
\vspace*{-1pc}
\label{mhprecision}
\end{table}

In Table~\ref{mhprecision}, the following assumptions have been made for
the various collider runs shown. TeV-33 results assume a total
integrated luminosity of $L=30\protect\fbi$.
LHC results assume $L=600\fbi$, which
corresponds to running two detectors
(ATLAS and CMS) for three years at LHC design luminosity.
Three NLC scenarios are listed corresponding
to three choices of center-of-mass energy:
(i) $\rts=500\gev$, (ii)~%
$\protect{\rts=\sqrt{s_{Zh}}\equiv \mz+\mhsm+20\gev}$, and
(iii) $\rts\simeq\mz+\mhsm$ ({\it i.e.}, threshold for $e^+e^-\to Z\hsm$).
In cases (i) and (ii), we assume $L=200\fbi$ and employ
the best tracking/calorimetry scenario outlined in
Ref.~\protect\cite{gunreport}.  NLC threshold results [case (iii)]
assume $L=50\fbi$ and are quoted {\it before} initial state radiation
and beam energy smearing effects are included.
In the latter case, including such effects would increase the
quoted errors by about 35\%.  The NLC results are also applicable
to the FMC, although with a 15\% increase in error in the last case if
all the cited effects were included.  Finally, the most accurate mass
measurements can be obtained by a scan at the FMC for the $s$-channel
Higgs resonance.  The FMC scan results listed in Table~\ref{mhprecision}
assume that a total luminosity of $L=200\fbi$ is devoted to the scan.

Precision measurements of heavy Higgs masses may also play an important
role in the study of Higgs phenomena.
In the decoupling limit, these mass splittings are of ${\cal
O}(\mz^2/\mha)$, which presents a formidable challenge to the design of
future Higgs searches.  Here is one case where the mass resolution
offered by the FMC might be required.  For example, it may be possible
to resolve the two peaks in a resonance scan for $\mu^+\mu^-\to\hh,
\ha$.  A measurement of the corresponding mass difference of the two
states would probe the structure of the electroweak symmetry breaking
dynamics.

Precision measurements of Higgs properties also include
branching ratios, cross-sections, and quantum numbers as discussed in
Phase 2 above.  One must be able to separate cross-sections and
branching ratios (instead of simply measuring the product of the two).
More
challenging will be the measurement of absolute partial widths, which
requires a determination of the total Higgs width.
Below $ZZ$ threshold, the Standard Model Higgs width is too small to be
directly measured, and other strategies must be employed.\footnote{The
width of $\hsm$ can be measured directly via $\hsm\to
ZZ\to\ell^+\ell^-\ell^+\ell^-$, if $\mhsm\gsim$ 190 GeV.
However, in models of non-minimal Higgs sectors, the mass of
the Higgs scalar with appreciable couplings to $ZZ$ typically lies below
this bound.}  As an illustration, Table~\ref{widthbrs} presents the
anticipated errors in the measurements of some $\hsm$ branching
ratios, the partial decay rate for $\hsm\to\gamma\gamma$, and the total
Higgs width, $\gamhsm$, for $80\leq\mhsm\leq
300\gev$.  The quoted errors are determined primarily
by considering the data that would be collected by the NLC at
$\rts=500\gev$ with a total integrated luminosity of $L=200\fbi$.
For BR($\hsm\to\gam\gam$), the NLC analysis has been combined
with results from an LHC analysis; while the measurement of
$\Gamma(\hsm\to\gam\gam)$ relies on data taken from a 50
fb$^{-1}$ run in the $\gamma\gamma$ collider mode of the NLC
(with the corresponding
$e^+e^-$ center-of-mass energy of $\sqrt{s}\sim 1.2\mhsm$).
These quantities also contribute to the net accuracy of the
total Higgs width, $\gamhsm$, following the indirect
procedure\footnote{For $\mhsm\lsim 130$~GeV, the indirect procedure
relies on the $\hsm\to\gamma\gamma$ measurements.  For
$\mhsm\gsim 130$~GeV, one may also
make use of the $WW\hsm$ coupling strength extracted from data.}
discussed in Ref.~\cite{gunreport}.
Note that $\gamhsm$ can be measured directly
only in the $s$-channel Higgs production at the FMC.  For comparison
with the indirect determination of $\gamhsm$,
the FMC scan results listed in Table~\ref{widthbrs}
assume that a total luminosity of $L=200\fbi$ is devoted to the scan.
With the exception of the case where $\mhsm\simeq\mz$, the FMC would
provide the most precise measurement of the total Higgs width for
values of the Higgs mass below the $W^+W^-$ threshold.

\begin{table}[h]
\caption{Anticipated experimental errors in the measured values of the
$\hsm$  branching ratios, the partial decay rate, $\Gamma(\hsm\to\gamma
\gamma)$, and total width, $\gamhsm$, in percent, for
various ranges of $\mhsm$.  The notation ``?'' indicates that a
reliable simulation or estimate is not yet available or that the number
indicated is a very rough guess, while ``--''
means that the corresponding observable cannot be reliably measured.
The results listed below are primarily derived from a multi-year run at
the NLC.  For $\hsm\to\gamma\gamma$, data from LHC and the
$\gamma\gamma$ collider are also employed to improve the quoted errors.
The total Higgs decay rate can be obtained indirectly (by combining
measurements of related quantities); the comparison with the direct
determination via $s$-channel Higgs resonance production at the FMC is
shown.  See the text and Ref.~\protect\cite{gunreport} for further
details.}
\renewcommand\arraystretch{1.5}
\setlength{\tabcolsep}{4.5pt}
\begin{center}
\begin{tabular}{@{}lcccc@{}}
\hline
\hline
 &  \multicolumn{4}{c}{$\mhsm$ range (GeV)} \\[-3pt]
\multicolumn{1}{c}{Observable}&
              80--130 &  130--150 &  150--170 & 170--300\\[3pt]
\hline \noalign{\vskip3pt}
 $\br(\hsm\to b\anti b)$& 5--6\% & 6--9\% & 20\% ? & $-$ \\
 $\br(\hsm\to c\anti c)$& $\sim 9\%$ & ? & ? & $-$ \\
 $\br(\hsm\to W\wstar)$\hspace*{-9pt} & $-$ & 16--6\% &6--5\% & 5--14\%
\\
 $\br(\hsm\to\gam\gam)$ & 15\% & 20--40\% & ? & $-$  \\
 $\Gamma(\hsm\to\gam\gam)$ & 12--15\% & 15--31\% & ? & 13--22\% \\
 $\gamhsm$ (indirect) & 19--13\% & 13--10\% & 10--11\% &  11--28\% \\
 $\gamhsm$ (FMC) & 3\%$^a$& 4--7\%& $-$& $-$\\[3pt]
\hline \hline
\end{tabular}
\end{center}
\vskip-3pt
\noindent
$^a$Near the $Z$ peak, the expected FMC uncertainty in $\gamhsm$
 is about 30\%.
\label{widthbrs}
\end{table}

In models of non-minimal Higgs sectors, precision measurements of the
branching ratios and partial (and total) decay rates of the lightest
CP-even Higgs boson could prove that $\hl\neq\hsm$, thereby
providing indirect evidence of the non-minimal Higgs states.
Once the non-minimal Higgs bosons are directly discovered, detailed
measurements
of their properties would yield significant clues to the underlying
structure of electroweak symmetry breaking.
For example, if the Higgs sector arises from a two-doublet model,
then precision studies of the heavy Higgs states can provide a
direct measurement of the important parameter $\tanb$ (the ratio of
Higgs vacuum expectation values).\footnote{Note that in the decoupling
limit (where $\hl$ cannot be distinguished from $\hsm$),
measurements of processes involving
$\hl$ alone cannot yield any information on the value of $\tanb$.}
The measurement of $\tanb$ can also provide a critical
self-consistency test of the MSSM, since the parameter $\tanb$ also
governs the properties of the charginos and neutralinos (and can in
principle be determined in precision measurements of supersymmetric
processes).  Moreover, the couplings of Higgs bosons to supersymmetric
particles will provide invaluable insights into both the physics of
electroweak
symmetry breaking and the structure of low-energy supersymmetry.  The
possibility that the heavy non-minimal Higgs states have non-negligible
branching ratios to supersymmetric partners can furnish an additional
experimental tool for probing the Higgs boson--supersymmetry connection.

As in the case of the
$\hsm$ discussed above, the lepton colliders (assuming $\sqrt{s}\gsim
2\mha$ for the NLC and $\sqrt{s}\sim\mha$ for the FMC) provide the
most powerful set of tools for extracting the magnitudes of the
Higgs couplings to fermion and vector boson pairs.  The Higgs couplings
to vector boson pairs directly probe the mechanism of electroweak
symmetry breaking [via the sum rule of eq.~(\ref{sumrule})].
The Higgs coupling to two photons, depends (through their one-loop
contributions)
on all charged states whose masses are generated by their
couplings to the Higgs sector.  Precision
measurements of the Higgs couplings to fermions are sensitive to other
Higgs sector parameters ({\it e.g.}, $\tan\beta$ and the neutral
Higgs mixing parameter $\alpha$ in a two-Higgs-doublet model).
Additional information can be ascertained if Higgs self-interactions
could be directly measured.  This would in principle provide direct
experimental access to the Higgs potential.  Unfortunately, there are
very few cases where the measurement of Higgs self-couplings has been
shown to be viable \cite{DHZ}.

Finally, one should also consider the possible effects of virtual Higgs
interactions \cite{VIRTUAL}.
In some models, flavor changing neutral currents mediated
by neutral Higgs bosons may be observable.  The CP-properties of the
heavy Higgs states could be mixed,\footnote{In the decoupling limit, the
lightest neutral scalar must be (approximately) a pure CP-even
state.}
leading to Higgs mediated
CP-violating effects that could be observed in processes with heavy
flavor.  In some cases, precision measurements of low-energy observables
can be quite sensitive to the heavy Higgs states.  The canonical example
is the process $b\to s\gamma$, which can be significantly enhanced
due to charged Higgs boson exchange.  If there are no other competing
non-Standard Model contributions (and this is a big {\it if}), then
present data excludes charged Higgs masses less than about
250 GeV \cite{joanne}.  Eventually,
when non-minimal Higgs states are directly probed, it is essential to
check for the consistency between their properties as determined
from direct observation and from their virtual effects.

\section{CONCLUSIONS}

This working group has examined the potential of a program of future
precision electroweak measurements \cite{PRECGROUP}
and the search for weakly-coupled
Higgs boson at future colliders \cite{gunreport}.
The goal of such a program is to
address the outstanding problem of elementary particle physics: what is
the origin of electroweak symmetry breaking and the nature of the
dynamics responsible for it?

The Higgs search will consist of three phases:
\begin{enumerate}
\item discovery of the Higgs signal,
\item verification of the Higgs interpretation, and
\item precision analysis of the Higgs sector properties.
\end{enumerate}
Improvements of
precision electroweak measurements can provide an important consistency
check of the Higgs interpretation of a Higgs signal, in the same way
that the LEP electroweak data provided support for the Tevatron
interpretation of the top-quark events.  Discovery of a Higgs-like signal
alone may not be sufficient to earn a place in the Particle Data Group
(PDG) tables.  Some basic measurements of the properties of the Higgs
candidate will be essential to confirm a Higgs interpretation of the
discovery.  Higgs searches at LEP-2, Run-II of the Tevatron and/or the
LHC will almost certainly discover a Higgs signal if electroweak
symmetry breaking dynamics is weakly coupled.  Moreover, measurements at
these machines will yield evidence for the Higgs interpretation that is
sufficient to pass PDG muster in some fraction of the Higgs parameter
space.  The NLC (and FMC) can discover or definitively rule out the
existence of a Higgs boson in the intermediate Higgs mass regime (the
mass region most problematical for the Higgs search at hadron
colliders).  Once a Higgs boson is discovered, the lepton colliders
would play a decisive role in the precision measurement of Higgs
sector properties.

It is not unlikely that the first Higgs state to be discovered will be
experimentally indistinguishable from the Standard Model Higgs boson.
This occurs in many theoretical models that exhibit the decoupling
of heavy scalar states.  In this decoupling limit,
the lightest Higgs state, $\hl$ is a neutral CP-even
scalar with properties nearly identical to the $\hsm$, while the other
Higgs bosons of the non-minimal Higgs sector are heavy (compared
to the $Z$) and are approximately mass-degenerate.  Thus, discovery of
$\hl\simeq\hsm$ may shed little light on the dynamics underlying
electroweak symmetry breaking.  It is then crucial to directly detect
and explore the properties of the non-minimal Higgs states.
In particular, precision measurements are critical
in order to distinguish
between $\hl$ and $\hsm$ and/or to map out the properties of the
non-minimal Higgs states.  To accomplish these goals, future colliders
of the highest energies and luminosities, considered in this report,
are essential.

We have entered a new era in Higgs phenomenology.  The methods
by which the first Higgs signal will be identified are well known and
have been studied in great detail.  However, the most outstanding
challenge facing the future Higgs searches lies in
identifying and exploring
in detail the properties of the non-minimal Higgs states.  A successful
exploration will have a profound effect on our understanding of
TeV-scale physics.



%
%

\begin{thebibliography}{99}
%
\bibitem{PRECISION}
{M.~Demarteau}, FERMILAB-Conf-96/354 (1996) [hep-ex/9611019],
invited talk given at the {\sl Meeting of the Division of Particles
and Fields}, Minneapolis, MN, 10--15 August 1996.
%
\bibitem{LEPEWWG}
{D. Abbaneo {\it et al.}} [LEP Electroweak Working Group] and
B. Schumm and D. Su [SLD Heavy Flavor Group], CERN-PPE/96-183 (1996),
prepared from contributions of the LEP and SLD
Experiments to the 28th International Conference on High Energy
Physics, Warsaw, Poland, 25--31 July 1996.
%
\bibitem{NEWPHENO}
See, {\it e.g.}, S. Godfrey, {\it et al.},
Summary of the New Phenomena Working Group,
in these Proceedings.
%
\bibitem{HUNTERS}
For a review of Higgs boson physics, see, {\it e.g.},
{J.F. Gunion, H.E. Haber, G.L. Kane and S. Dawson},
{\it The Higgs Hunter's Guide} (Addison-Wesley, Redwood City, CA, 1990).
%
\bibitem{SUSY} For reviews on low-energy supersymmetry,
see, {\it e.g.},
{H.E. Haber and G. Kane}, {\sl Phys. Rep.} {\bf 117}, 75 (1985);
H.E. Haber, in {\it Recent Directions in Particle Theory}, Proceedings
of the 1992 Theoretical Advanced Study Institute in Elementary Particle
Physics, edited by J. Harvey and J. Polchinski (World Scientific,
Singapore, 1993) pp.~589--686; R. Arnowitt and P. Nath, in {\sl
Particles and Fields}, Proceedings of the VII Jorge Andre
Swieca Summer School, Sao Paulo, Brazil, 10--23 January, 1993, edited by
J.P. Eboli and V.O. Rivelles (World Scientific, Singapore, 1994)
pp.~3--63.
%
\bibitem{SUSYGROUP} {J. Bagger {\it et al.}},
Summary of the Supersymmetry Working Group, in these Proceedings;
{J. Amundson {\it et al.}}, Report of the Supersymmetry Theory
Subgroup, in these Proceedings.
%
\bibitem{rosner}
See, {\it e.g.}, J. Rosner, CERN-TH/96-245 (1996) [hep-ph/9610222], to
appear in {\it Masses of Fundamental Particles},
Proceedings of the Carg\`ese Summer Institute on Particle Physics,
August 1996, to be published by Plenum Press.
%
\bibitem{TC}
For a review of theories of dynamical electroweak symmetry breaking,
see {\it e.g.}, K. Lane, in {\it The Building Blocks of Creation},
Proceedings of the 1993 Theoretical Advanced Study Institute in
Elementary Particle Physics, edited by S. Raby and T. Walker (World
Scientific, Singapore, 1994) pp.~381--408.
%
\bibitem{SEWS} {T.L. Barklow {\it et al.}},  Summary of the
Strong Coupling Electroweak Symmetry Breaking Working Group, in
these Proceedings.
%
\bibitem{nmssm}
J. Ellis, J.F. Gunion, H.E. Haber, L. Roszkowski and F. Zwirner,
{\sl Phys. Rev.} {\bf D39}, 844 (1989);
M. Drees, {\sl Int. J. Mod. Phys.} {\bf A4}, 3635 (1989);
U. Ellwanger, M. Rausch de Traubenberg and C.A. Savoy,
{\sl Phys.\ Lett.} {\bf B315}, 331 (1993); {\sl Z.\ Phys.}
{\bf C67}, 665 (1995); LPTHE-ORSAY-96-85 [hep-ph/9611251];
T. Elliot, S.F. King and P.L. White, {\sl Phys. Lett.} {\bf B305}, 71
(1993); {\sl Phys. Rev.} {\bf D49}, 2435 (1994); S.F. King and P.L.
White, {\sl Phys.\ Rev.} {\bf D52}, 4183 (1995).
%
\bibitem{rparity}
See, {\it e.g.}, F. de Campos, M.A. Garcia-Jareno, A.S. Joshipura,
J. Rosiek, and J.W.F. Valle, {\sl Nucl. Phys,} {\bf B451}, 3 (1995);
T. Banks, Y. Grossman, E. Nardi, and Y. Nir, {\sl Phys. Rev.}
{\bf D52}, 5319 (1995).
%
\bibitem{lepdesign}
S. Meyers and C. Wyss, in {\it Physics at LEP2}, Volume 1, edited by G.
Altarelli, T. Sj\"ostrand and F. Zwirner, CERN Yellow Report 96-01
(1996) pp.~23--43.
%
\bibitem{PRECGROUP} {U. Baur and M. Demarteau {\it et al.}},
Precision Electroweak Physics Subgroup Report, in these Proceedings.
%
\bibitem{gunreport}
J.F. Gunion {\it et al.}, Higgs Boson Discovery and Properties
Subgroup Report, in these Proceedings.
%
\bibitem{PT} {M. E. Peskin and T. Takeuchi},
{\sl Phys. Rev. Lett.} {\bf 65}, 964 (1990); {\sl Phys. Rev.}
{\bf D46}, 381 (1992).
%
\bibitem{LangAlt} {P. Langacker},
invited talk given at the {\sl Meeting of the Division of Particles
and Fields}, Minneapolis, MN, 10--15 August 1996;
G. Altarelli, CERN-TH-96-265 (1996) [hep-ph/9611239],
talk given at the NATO Advanced Study Institute on Techniques
and Concepts of High-Energy Physics, St. Croix, U.S. Virgin Islands,
July 10-23, 1996.
%
\bibitem{CCFR}
{K.S. McFarland} {\it et al.} (CCFR/NuTeV Collaboration),
in {\it Electroweak Interactions and Unified Theories},
Proceedings of the XXXI Rencontres de Moriond, Les Arcs, France,
16--23 March, 1996, edited by
J. Tr\^an Thanh V\^an (Editions Fronti\`eres, Gif-sur-Yvette, France, 1996)
pp.~97--102.
%
\bibitem{ecfa} J.-F. Grivaz, T. Sj\"ostrand and P.M. Zerwas, conveners
of the Physics Working Group, to appear in the Proceedings of the
Joint ECFA/DESY Study: Physics and Detectors for a Linear Collider,
February--November, 1996.
%
\bibitem{muonmwmt} V. Barger, M. Berger, J.F. Gunion and T. Han,
MADPH-96-963 and UCD-97-01 (1997) [hep-ph/9702334].
%
\bibitem{SLC2000}
M. Breidenbach {\it et al.}, SLAC-CN-409 (1996).
%
\bibitem{berndk}
B.A. Kniehl, {\sl Z. Phys.} {\bf C72}, 437 (1996).
%
\bibitem{scottw}
M.C. Smith and S.S. Willenbrock, Illinois preprint (1996)
[hep-ph/9612329].
%
\bibitem{habernir} {H. E. Haber and Y. Nir},
{\sl Nucl. Phys.} {\bf B335}, 363 (1990).
%
\bibitem{DECP}
{H.E. Haber}, in {\it Beyond the Standard Model IV},
Proceedings of the Fourth International Conference
on Physics Beyond the Standard Model, Granlibakken,
Lake Tahoe, CA, 13--18 December, 1994, edited by J.F. Gunion,
T. Han and J. Ohnemus (World Scientific, Singapore, 1995) pp.~151--163;
and in {\it Perspectives for Electroweak Interactions in $e^+e^-$
Collisions}, Proceedings of the Ringberg Workshop, Ringberg Castle,
Tegernsee, Germany, 5--8 February, 1995, edited by B.A. Kniehl
(World Scientific, Singapore, 1995) pp.~219--231.
%
\bibitem{lattice} For a recent review, see, {\it e.g.},
U.M. Heller, {\sl Nucl. Phys B. (Proc. Suppl.)} {\bf 34},
101 (1994).
%
\bibitem{cabibbo} N. Cabibbo, L. Maiani, G. Parisi and R. Petronzio,
{\sl Nucl. Phys.} {\bf B158}, 295 (1979); M. Lindner,
{\sl Z. Phys.} {\bf C31}, 295 (1986).
%
\bibitem{schrempp}
For a recent review, see B. Schrempp and M. Wimmer, {\sl Prog. Part.
Nucl. Phys.} {\bf 37}, 1 (1996).
%
\bibitem{altarelli}
{G. Altarelli and G. Isidori},
{\sl Phys. Lett.} {\bf B337}, 141 (1994).
%
\bibitem{quiros}
{J.A. Casas,
J.R. Espinosa and M. Quiros}, {\sl Phys. Lett.} {\bf B342}, 171 (1995);
{J.R. Espinosa and M. Quiros}, {\sl Phys. Lett.} {\bf B353}, 257 (1995).
%
%
%
\bibitem{hhg}
See, {\it e.g.}, Chapter 4 of Ref.~\cite{HUNTERS}.
%
\bibitem{hhprl}
{H.E. Haber and R. Hempfling}, {\sl Phys. Rev. Lett.}
{\bf 66}, 1815 (1991);
{Y. Okada, M. Yamaguchi and T. Yanagida},
{\sl Prog. Theor. Phys.} {\bf 85}, 1 (1991);
{J. Ellis, G. Ridolfi and F. Zwirner},
{\sl Phys. Lett.} {\bf B257}, 83 (1991).
%
\bibitem{honeloop}
{P. Chankowski, S. Pokorski and J. Rosiek}, {\sl Phys. Lett.} {\bf
B274}, 191 (1992); {\sl Nucl. Phys.} {\bf B423}, 437 (1994);
{A. Dabelstein}, {\sl Z. Phys.} {\bf C67}, 495 (1995);
{D.M. Pierce, J.A. Bagger, K. Matchev and R.-J. Zhang},
SLAC-PUB-7180 (1996) [hep-ph/9606211], {\sl Nucl. Phys. B} (1997), in press.
%
\bibitem{llog}
{H.E. Haber and R. Hempfling}, {\sl Phys. Rev.} {\bf D48}, 4280 (1993).
%
\bibitem{hempfhoang}
{R. Hempfling and A.H. Hoang}, {\sl Phys. Lett.} {\bf B331}, 99 (1994).
%
\bibitem{hhh}
{H.E. Haber, R. Hempfling and A.H. Hoang}, CERN-TH/95-216 (1996)
[hep-ph/9609331], {\sl Z. Phys. C} (1997), in press.
%
\bibitem{carena}
{M. Carena, J.R. Espinosa, M. Quiros and C.E.M. Wagner},
{\sl Phys. Lett.} {\bf B355}, 209 (1995);
{M. Carena, M. Quiros and C.E.M. Wagner},
{\sl Nucl. Phys.} {\bf B461}, 407 (1996).
%
\bibitem{GKW}
{G. L. Kane, C. Kolda and J. D. Wells},
{\sl Phys. Rev. Lett.} {\bf 70}, 2686 (1993).
%
%
\bibitem{ypan}
Y. Pan [for the ALEPH Collaboration], invited talk given at the
International Symposium on Recent Developments in Phenomenology,
Madison, Wisconsin, 17--19 March 1997.
%
\bibitem{LEPHIGGS}
J.P. Martin, LYCEN-9644 (1996), invited talk given at the 28th
International
Conference on High Energy Physics, Warsaw, Poland, 25--31 July 1996.
%
%
\bibitem{CDFHIGGS}
{C. Loomis}, CDF Collaboration,
FERMILAB-CONF-96-232-E (1996), talk presented at the DPF-96 Conference,
Minneapolis, MN, 10--15 August 1996.
%
\bibitem{janot}
M. Carena, P.M. Zerwas {\it et al.}, in {\it Physics at LEP2},
Volume 1, edited by G. Altarelli, T. Sj\"ostrand and F. Zwirner,
CERN Yellow Report 96-01 (1996) pp.~351--462.
%
\bibitem{tevreport}
D. Amidei and R. Brock, editors, {\it Future Electroweak Physics at the
Fermilab Tevatron}, Report of the TeV-2000 Study Group,
FERMILAB-Pub-96/082 (1996).
%
\bibitem{kky}
S. Kim, S. Kuhlmann and W.-M. Yao, ``Improvement of signal significance
in $WH\to\ell\nu b\bar b$ search at TeV-33'', CDF-ANAL-EXOTIC-PUBLIC-3904
(1996), contribution to these Proceedings.
%
\bibitem{yao}
W.-M. Yao, ``Prospects for observing the Higgs boson in
the $ZH\to(\nu\bar\nu,\ell^+\ell^-)b\bar b$ channel at TeV-33'',
FERMILAB-CONF-96-383-E (1996), contribution to these Proceedings.
%
\bibitem{atlas}
W.W. Armstrong {\it et al.} [ATLAS Collaboration], {\it ATLAS Technical
Proposal}, CERN-LHCC-94-43 (1994).
%
\bibitem{cms}
G.L. Bayatian {\it et al.} [CMS Collaboration], {\it CMS Technical
Proposal}, CERN-LHCC-94-38 (1994).
%
\bibitem{Dreiner}
E.W.N. Glover, J. Ohnemus and S.S.D Willenbrock,
{\sl Phys. Rev.} {\bf D37}, 3193 (1988);
V. Barger, G. Bhattacharya, T. Han and B.A. Kniehl,
{\sl Phys. Rev.} {\bf D43}, 779 (1991);
M. Dittmar and H. Dreiner, {\sl Phys.\ Rev.} {\bf D55}, 167 (1997).
%
\bibitem{finland}
H.E. Haber, in {\it Physics and Experiments with Linear Colliders},
Volume I, Proceedings of the Linear Collider Workshop, Saariselk\"a,
Finland, 1--14 September 1991, edited by R. Orava, P. Eerola, and M.
Nordberg (World Scientific, Singapore, 1992) pp.~235--275.
%
\bibitem{nlcreport}
S. Kuhlman {\it et al.}, {\it Physics and Technology of the Next Linear
Collider}, a report submitted to Snowmass '96, SLAC-Report-485 (1996).
%
\bibitem{desyreport}
A. Djouadi {et al.}, ``Higgs Physics'' [hep-ph/9605437], prepared by the
European Higgs working group on Physics with $e^+e^-$ Linear Colliders.
%
\bibitem{BBGH}
{V. Berger, M. Berger, J.F. Gunion and T. Han}, {\sl Phys.
Rev. Lett.} {\bf 75}, 1462 (1995); UCD-96-6 (1996) [hep-ph/9602415],
to appear in {\sl Phys. Rep.}
%
\bibitem{mumureport}
R. Palmer {\it et al.} [The $\mu^+\mu^-$ Collider Collaboration], {\it
$\mu^+\mu^-$ Collider: A Feasibility Study}, BNL-52503 (1996).
%
\bibitem{miller}
D.J. Miller, in Proceedings of the Second Workshop on the Physics
Potential and Development of $\mu^+\mu^-$ Colliders, Sausalito, CA,
1994, edited by D. Cline (American Institute of Physics Conference
Proceedings 352) p.~191.
%
\bibitem{ghw}
{J.F. Gunion, H.E. Haber and J. Wudka}, {\sl Phys. Rev.}
{\bf D43}, 904 (1991).
%
\bibitem{KOT}
{J. Kamoshita, Y. Okada and M. Tanaka},
{\sl Phys.\ Lett.} {\bf B328}, 67 (1994).
%
\bibitem{GHM}
{J.F. Gunion, H.E. Haber and T. Moroi}, UCD-96-26 (1996)
[hep-ph/9610337], contribution to these Proceedings.
%
\bibitem{gunhab}
{J.F. Gunion and H.E. Haber}, {\sl Phys. Rev.} {\bf D48},
5109 (1993); D.L. Borden, D.A. Bauer and D.O. Caldwell,
{\sl Phys.\ Rev.} {\bf D48}, 4018 (1993).
\vfill\eject
%
\bibitem{bctags}
R. Van Kooten and D. Jackson, work reported in Ref.~\cite{gunreport}
(see Fig.~3).
%
\bibitem{luc}
{L. Poggioli}, private communication and ATLAS technical note in
preparation.
%
\bibitem{patrick}
P. Janot, private communication (analysis based on the work of
Ref.~\cite{janot}).
%
\bibitem{DHZ}
A. Djouadi, H.E. Haber and P.M. Zerwas, {\sl Phys. Lett.}
{\bf B375}, 203 (1996); F. Boudjema and E. Chopin, {\sl Z. Phys.} {\bf
C73}, 85 (1996).
%
\bibitem{VIRTUAL}
J.L. Hewett, T. Takeuchi and S. Thomas, in
{\it Electroweak Symmetry Breaking and New Physics at the TeV Scale},
edited by T.L. Barklow, S. Dawson, H.E. Haber, and J.L. Siegrist
(World Scientific, Singapore, 1996), pp.~548-649.
%
\bibitem{joanne}
J.L. Hewett, {\sl Phys. Rev. Lett.} {\bf 70}, 1045 (1993);
V. Barger, M. Berger and R.J.N. Phillips, {\sl Phys. Rev. Lett.} {\bf
70}, 1368 (1993); J.L. Hewett, in {\it B Physics: Physics Beyond
the Standard Model at the B Factory}, Proceedings of the International
Workshop on $B$ Physics, Nagoya, Japan, 26--28 October 1994, edited by
A.I. Sanda and S. Suzuki (World Scientific, Singapore, 1995)
pp.~321--326.
%
\end{thebibliography}
\end{document}